\documentclass[epsfig,usenatbib]{mnras}
\usepackage{epsfig}
\usepackage{natbib}
\usepackage{amsmath}

\def\gtsima{$\; \buildrel > \over \sim \;$}
\def\ltsima{$\; \buildrel < \over \sim \;$}
\def\gtrsim{\lower.5ex\hbox{\gtsima}}
\def\lesssim{\lower.5ex\hbox{\ltsima}}

\begin{document}
\title[The properties of merging BHs and NSs]{The properties of merging black holes and neutron stars across cosmic time}
\author[Mapelli et al.]
{Michela Mapelli$^{1,2,3,4}$, Nicola Giacobbo$^{1,2,3}$, Filippo Santoliquido$^{1}$, M. Celeste Artale$^{4}$ 
\\
%$^1$Institute for Astrophysics and Particle Physics, University of Innsbruck, Technikerstrasse 25/8, 6020, Innsbruck, Austria\\
$^1$Physics and Astronomy Department Galileo Galilei, University of Padova, Vicolo dell'Osservatorio 3, I--35122, Padova, Italy, \\{\tt michela.mapelli@unipd.it}\\
$^2$INAF-Osservatorio Astronomico di Padova, Vicolo dell'Osservatorio 5, I--35122, Padova, Italy \\
$^3$INFN-Padova, Via Marzolo 8, I--35131 Padova, Italy\\
$^4$Institut f\"ur Astro- und Teilchenphysik, Universit\"at Innsbruck, Technikerstrasse 25/8, A--6020, Innsbruck, Austria\\
}
\maketitle \vspace {7cm }
\bibliographystyle{mnras}
 
\begin{abstract}
  The next generation ground-based gravitational wave interferometers will possibly observe mergers of binary black holes (BBHs) and  binary neutron stars (BNSs) to redshift $z\gtrsim{}10$ and $z\gtrsim{}2$, respectively. Here, we characterize the properties of merging BBHs, BNSs and neutron star-black hole binaries across cosmic time, by means of population-synthesis simulations combined with the Illustris cosmological simulation. We find that the  mass of merging compact objects does not depend (or depends very mildly) on the merger redshift. Even the mass distribution of black holes depends only mildly on redshift, because BBHs originating from metal-poor progenitors ($Z\leq{}4\times{}10^{-3}$) dominate the entire population of merging BBHs across cosmic time.  For a common-envelope efficiency $\alpha{}\ge{}3$, the main difference between the mass distribution of BBHs merging in the last Gyr and that of BBHs merging more than 11 Gyr ago is that there is an excess of heavy merging black holes ($20-35$ M$_\odot$) in the last Gyr. This excess is explained by the longer delay time of massive BBHs. 
\end{abstract}

%
%________________________________________________________________

\section{Introduction}
The first two observing runs of the advanced LIGO \citep{LIGOdetector} and Virgo interferometers \citep{VIRGOdetector} have led to the detection of ten binary black hole (BBH) mergers \citep{abbottGW150914,abbottGW151226,abbottO1,abbottGW170104,abbottGW170608,abbottO2} and one binary neutron star (BNS) merger \citep{abbottGW170817,abbottmultimessenger}. From these detections, we can attempt to reconstruct the properties of BBHs merging in the local Universe: %the mass distribution of black holes (BHs) is consistent
the black holes (BHs) detected thus far are consistent with a power law mass distribution with index $1.6^{+1.5}_{-1.7}$ (at 90\% credibility) and with maximum mass $\sim{}45$ M$_\odot$ (e.g. \citealt{abbottO2popandrate}). The third observing run of LIGO-Virgo (O3), which will start in few months, is expected to bring tens of new detections, improving significantly our knowledge of the mass distribution of BBH mergers in the local Universe.  

Third-generation ground-based gravitational wave (GW) detectors are now being planned \citep{punturo2010}. The European Einstein Telescope and its American peer Cosmic Explorer will detect BBH mergers across the entire Universe and will probe BNS mergers up to redshift $z\gtrsim{}2$, which is the peak of cosmic star formation rate \citep{madaudickinson2014}. Thus, it is of crucial importance to investigate the redshift evolution of merging compact binaries, in preparation for O3 and especially for third-generation ground-based GW detectors. In particular, the predicted dependence of BH mass on the metallicity of the stellar progenitors \citep{heger2003,mapelli2009,mapelli2010,mapelli2013,belczynski2010,fryer2012,spera2015} might suggest that more massive BHs form in the higher redshift Universe, where the average metallicity was lower.

 Several studies have investigated the cosmic evolution of the merger rate density (e.g. \citealt{oshaughnessy2010,dominik2013,dominik2015,belczynski2016,dvorkin2016,demink2016,mandel2016,elbert2017,mapelli2017,mapelli2018,eldridge2019,fishbach2018,rodriguez2018}), suggesting that the merger rate of BBHs, neutron star-black hole binaries (NSBHs) and BNSs increases with redshift, reaching a peak at $z\sim{}2-4$. Other works have tried to characterize the host galaxies of merging compact objects (e.g. \citealt{oshaughnessy2010,lamberts2016,oshaughnessy2017,schneider2017,cao2018,mapelli2018b,lamberts2018,marassi2019,artale2019}). However, none of the previous studies focuses on the evolution of the population of merging compact objects across cosmic time. Is the typical mass of two BHs merging in the local Universe the same of two BHs merging at redshift $z\sim{}7$?

This paper studies the evolution of merging compact objects (BBHs, NSBHs and BNSs) across cosmic time. We investigate the redshift evolution of the mass spectrum of merging compact objects, of the metallicity of their progenitors and of the delay time (i.e. the time elapsed from the formation of the stellar progenitors and the merger of two compact objects). We use population-synthesis simulations \citep{giacobbo2018,giacobbo2018b} combined with the outputs of the Illustris cosmological simulation \citep{vogelsberger2014a,vogelsberger2014b} through a Monte Carlo procedure \citep{mapelli2017,mapelli2018}. This methodology associates merging compact objects to a given galaxy \citep{oshaughnessy2010,kelley2010,lamberts2016,oshaughnessy2017,schneider2017} based on its star formation rate and on the metallicity of its stellar particles.  Remarkably, we find that the mass distribution of merging compact objects depends only mildly on merger redshift.

\section{Methods}\label{sec:methods}
We present the results of binary population-synthesis simulations run with the code \textsc{mobse} convolved with the outputs of the \textsc{Illustris-1} cosmological simulation by means of a simple Monte Carlo algorithm. The methodology has already been described in \cite{mapelli2017} and \cite{mapelli2018}. Thus, herebelow we briefly summarize the main steps of our methodology and we refer to the aforementioned papers for more details. 
\subsection{Population-synthesis simulations with {\sc mobse}}\label{sec:section2.1}
\textsc{mobse} \citep{giacobbo2018,giacobbo2018b} is an upgraded version of the \textsc{bse} code \citep{hurley2000,hurley2002}. The main updates concern mass loss by stellar winds of massive hot stars, the mass of compact remnants and the magnitude of natal kicks. 

Mass loss of massive hot stars $\dot{M}$ is assumed to depend on both the metallicity $Z$ and the Eddington ratio $\Gamma_e=L_\ast{}/L_{\rm Edd}$ (where $L_\ast$ and $L_{\rm Edd}$ are the stellar luminosity and its Eddington value, respectively). In particular, we describe mass loss as $\dot{M}\propto{}Z^{\beta{}}$, where $\beta=0.85$ if $\Gamma{}_e<2/3$, $\beta=2.45-2.4\,{}\Gamma_e$ if $1>\Gamma{}_e\ge{}2/3$ and $\beta{}=0.05$ if $\Gamma_e\ge{}1$ (\citealt{chen2015}, see also  \citealt{vink2001,vink2005,graefener2008,vink2011}).

The mass of compact objects which form via core-collapse supernova (SN) is derived following \cite{fryer2012}. In particular, we assume that the final mass of the compact object depends on the final mass of the Carbon-Oxygen core and on the final total mass of the star. If the final mass of the Carbon-Oxygen core is $m_{\rm CO}\geq{}11$ M$_\odot$ the star is assumed to directly collapse into a black hole (BH) without SN explosion. In this paper, we adopt the rapid model presented in \cite{fryer2012}, which enforces the possible mass gap between the more massive neutron stars (NSs, $m_{\rm NS}\leq{}2$ M$_\odot$) and the lighter BHs ($m_{\rm BH}\geq{}5$ M$_\odot$, see \citealt{ozel2010,farr2011,kreidberg2012,littenberg2015}). The outcomes of electron-capture SNe are also included in \textsc{mobse}, as described in \cite{giacobbo2019}.

Finally, pair instability and pulsational pair instability SNe are implemented in \textsc{mobse} following \cite{spera2017}, as described in \cite{giacobbo2018}. If the Helium core of a star grows larger than $\sim{}64$ M$_\odot$, the star is completely destroyed by pair instability and leaves no compact object. In contrast, if the star has a Helium core mass $32\leq{}m_{\rm He}/m_\odot\leq{}64$, it undergoes pulsational pair instability, which leads to significant mass loss and to a smaller BH mass.

The stellar wind model combined with these prescriptions for SNe produces a mass spectrum of compact objects which depends on the metallicity of the progenitor star (see e.g. Figure~4 of \citealt{giacobbo2018}). BHs with mass up to $\sim{}65$ M$_\odot$ are allowed to form at low metallicity. 

Natal kicks are another critical ingredient of binary population-synthesis codes. In \textsc{mobse} we draw natal kicks from a Maxwellian distribution with one-dimensional root-mean-square velocity $\sigma{}$. For electron-capture SNe, we assume $\sigma{}_{\rm ECSN}=15$ km s$^{-1}$ (see the discussion in \citealt{giacobbo2019}). For core-collapse SNe, we adopt $\sigma{}_{\rm CCSN}=265$ km s$^{-1}$ in run~$\alpha{}5$ and $\sigma{}_{\rm CCSN}=15$ km s$^{-1}$ in run~CC15$\alpha{}5$. The former value is inferred by \cite{hobbs2005} from the analysis of the proper motion of $\sim{}230$ Galactic pulsars. The latter value was introduced by \cite{giacobbo2018b} to account for the small kicks associated with ultra-stripped SNe \citep{tauris2015,tauris2017} and for the possible evidence of small kicks in Galactic BNSs \citep{beniamini2016}.

Moreover, we take into account the effect of fallback by reducing the kick velocity as $\tilde{v}_{\rm kick}=(1-f_{\rm fb})\,{}v_{\rm kick}$, where $v_{\rm kick}$ is the kick extracted from the Maxwellian distribution, while $f_{\rm fb}$ is the fraction of mass which falls back onto the proto-neutron star (see \citealt{fryer2012}). If a BH forms via direct collapse $f_{\rm fb}=1$. Thus, BHs formed via direct collapse undergo no kick, apart from the Blaauw kick \citep{blaauw1961,boersma1961} associated with symmetric neutrino loss\footnote{In the case of a direct collapse, we assume that the gravitational remnant mass is $m_{\rm rem,\,{}grav}=0.9\,{}m_{\rm f}$, where $m_{\rm f}$ is the stellar mass at the onset of collapse, \citep{fryer2012}. Thus, even the Blaauw kick is generally negligible for direct-collapse BHs.}.

The main processes of binary evolution (wind mass transfer, Roche-lobe mass transfer, common envelope and tidal evolution) are implemented in \textsc{mobse} as described in \textsc{bse} \citep{hurley2002}. In particular, our treatment of common envelope (CE) depends on two parameters: $\alpha{}$ (describing the efficiency of energy transfer) and $\lambda{}$ (describing the geometry of the envelope and the importance of recombinations). In the current paper, $\lambda{}$ is defined as in \cite{claeys2014}, to account for the contribution of recombinations, while $\alpha$ is a constant. We adopt $\alpha{}=5$.  The main change in the description of CE with respect to {\sc bse} consists in the treatment of Hertzsprung gap (HG) stars. In the standard version of \textsc{bse},  HG donors entering a CE phase are allowed to survive the CE phase. In the version of {\sc{mobse}} we use here, HG donors are forced to merge with their companion if they enter a CE. In fact, {\sc{mobse}} simulations in which HG donors are allowed to survive a CE phase produce a local BBH merger rate $R_{\rm BBH}\sim{}600-800$ Gpc$^{-3}$ yr$^{-1}$, which is not consistent with LIGO-Virgo results \citep{mapelli2017}.

Decay by gravitational-wave emission is described as in \cite{peters1964}. In contrast to \textsc{bse}, we account for gravitational-wave decay for all compact-object binaries, even with semi-major axis $a>10$ R$_\odot$.
 %%%%%%%%%%%%%%%%%%%%%%%%%%%%%%%%%% TABLE 1%%%%%%%%%%%%%%%%%%%%%%%%%%%%%%%%%%%%%
\begin{table}
\begin{center}
\caption{\label{tab:table1}
Properties of the population-synthesis simulations.}
 \leavevmode
\begin{tabular}[!h]{llll}
\hline
Run    & $\alpha{}$ & $\sigma_{\rm ECSN}$ & $\sigma_{\rm CCSN}$\\
& & [km s$^{-1}$] & [km s$^{-1}$]\\
\hline
$\alpha5$     & 5.0        & 15 & 265  \\
CC15$\alpha5$ & 5.0        & 15 & 15 \\
\hline
\end{tabular}
\begin{flushleft}
\footnotesize{Column 1: model name; column 2: value of $\alpha{}$ in the CE formalism; column 3 and 4: one-dimensional root-mean square of the Maxwellian distribution for electron-capture SN kicks ($\sigma_{\rm ECSN}$) and core-collapse SN kicks ($\sigma_{\rm CCSN}$). See Section~\ref{sec:section2.1} for details. }
\end{flushleft}
\end{center}
\end{table}
%%%%%%%%%%%%%%%%%%%%%%%%%%%%%%%%%%%%%%%%%%%%%%%%%%%%%%%%%%%%%%%%%%%%%%%%%%%%%%%%
%%%%%%%%%%%%%%%%%%%%%%%%%%%%%%%%%%% FIGURE 1 %%%%%%%%%%%%%%%%%%%%%%%%%%%%%%%%%%
\begin{figure*}
\center{{
    \epsfig{figure=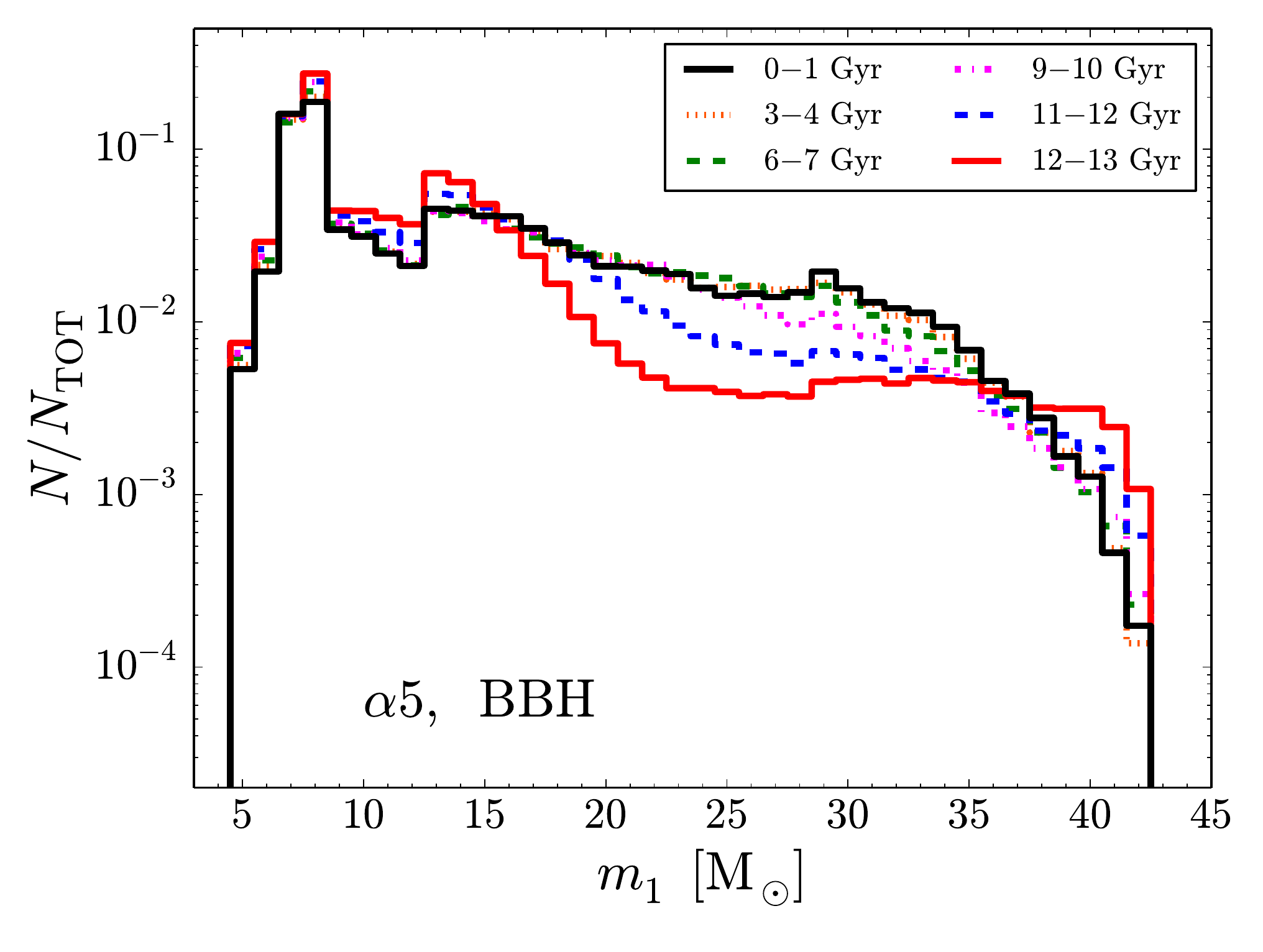,width=7cm} %was fig1.pdf
        \epsfig{figure=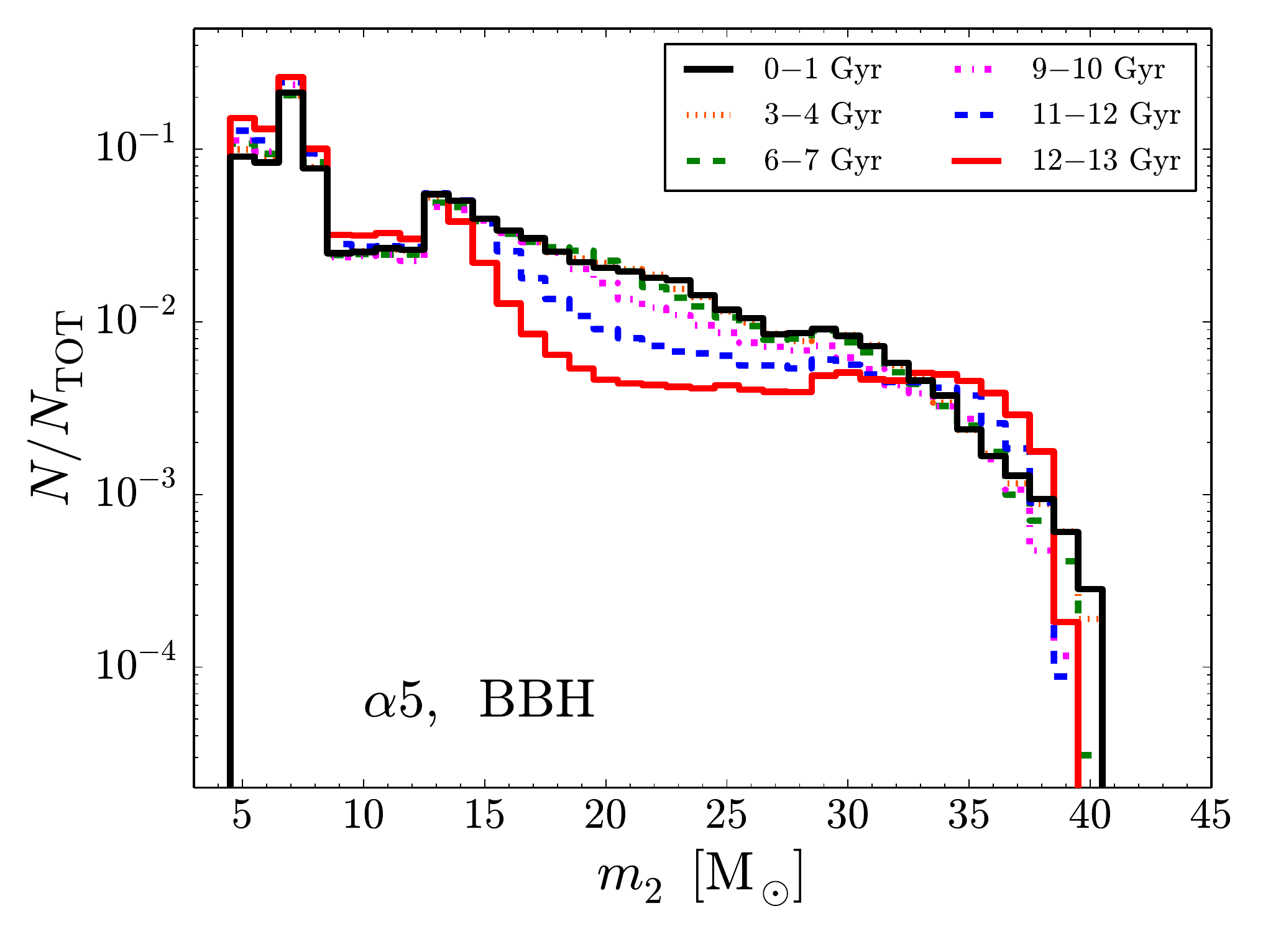,width=7cm} %was fig1.pdf
    \epsfig{figure=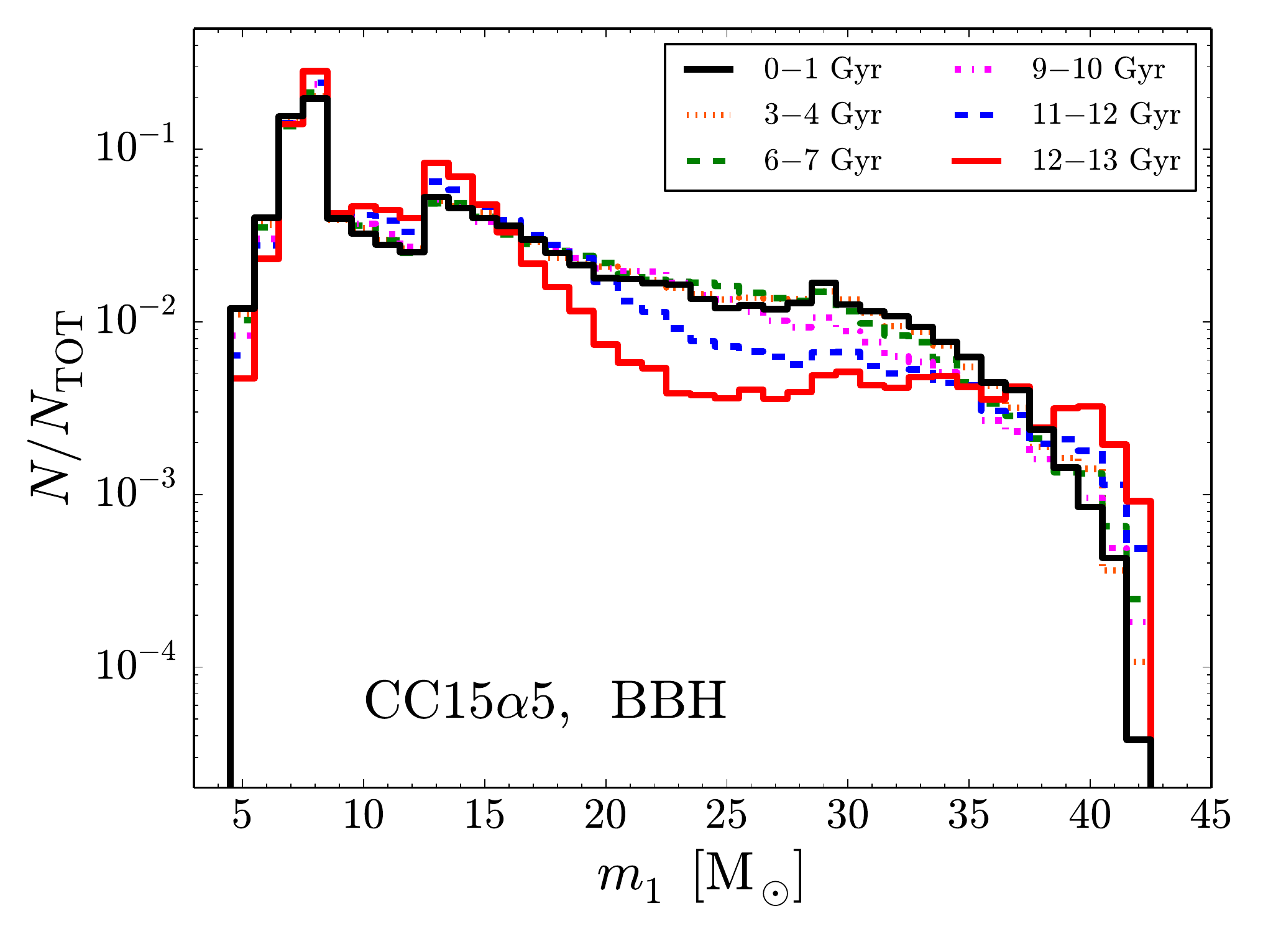,width=7cm} %was fig1.pdf
    \epsfig{figure=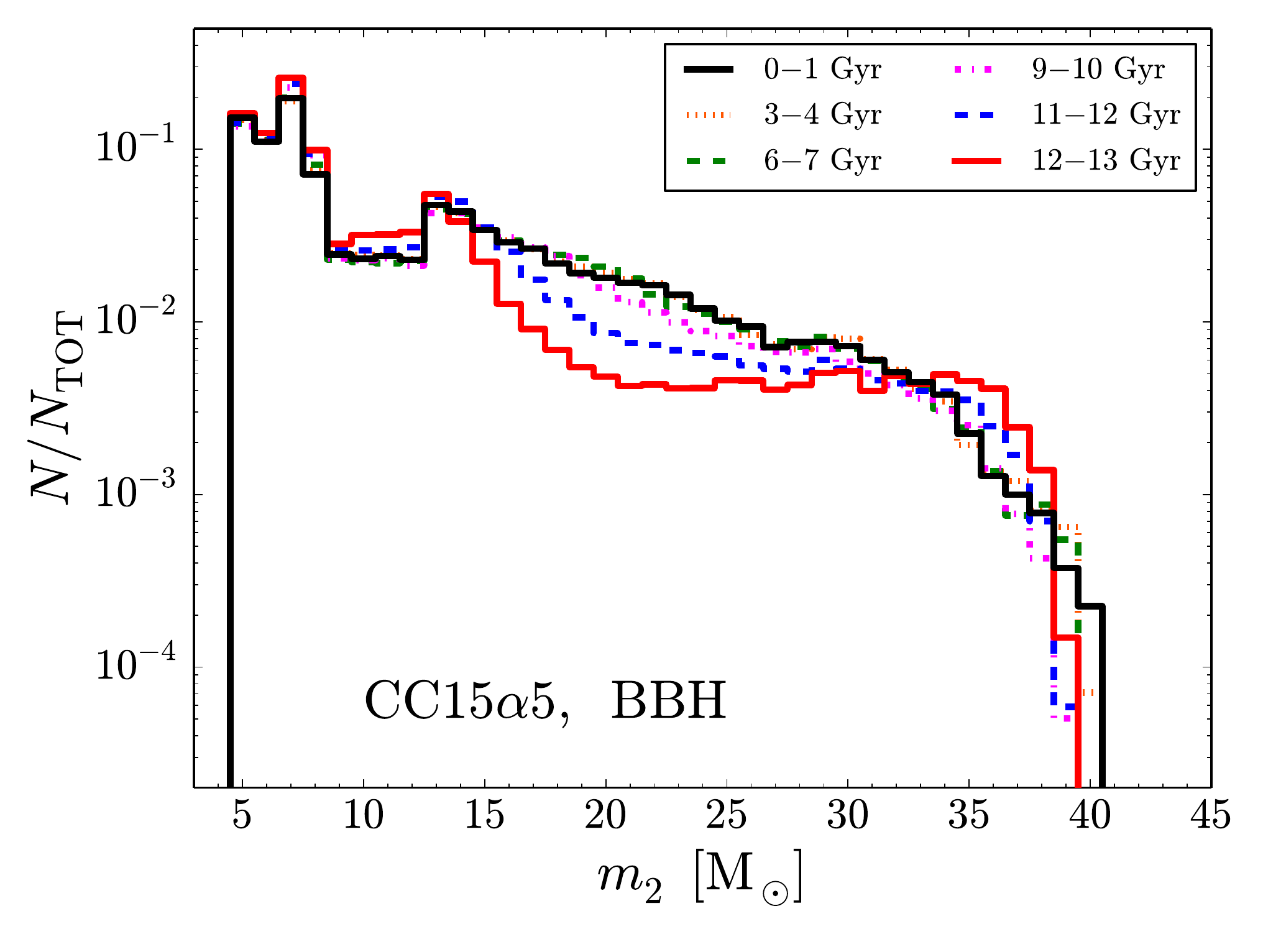,width=7cm} %was fig1.pdf
}}
    \caption{Mass of the primary BH ($m_1$, left-hand panels) and of the secondary BH ($m_2$, right-hand panel) in merging BBHs. Black solid line: BBHs merging $0-1$ Gyr ago ($z\leq{}0.08$); orange dotted line: BBHs merging $3-4$ Gyr ago ($z=0.26-0.37$); green dashed line: BBHs merging $6-7$ Gyr ago ($z=0.64-0.82$); magenta dot-dashed line:  BBHs merging $9-10$ Gyr ago ($z=1.35-1.78$); blue dashed line: BBHs merging $11-12$ Gyr ago ($z=2.43-3.65$); red solid line: BBHs merging $12-13$ Gyr ago ($z=3.65-7.15$). Top panels: run $\alpha{}5$. Bottom panels: run CC15$\alpha{}5$. The number of BHs $N$ on the $y-$axis is normalized to the total number of BHs $N_{\rm TOT}$ in each histogram.
      \label{fig:BBHmass}
}
\end{figure*}
%%%%%%%%%%%%%%%%%%%%%%%%%%%%%%%%%%%%%%%%%%%%%%%%%%%%%%%%%%%%%%%%%%%%%%%%%%%%%%%

%%%%%%%%%%%%%%%%%%%%%%%%%%%%%%%%%%% FIGURE 2 %%%%%%%%%%%%%%%%%%%%%%%%%%%%%%%%%%
\begin{figure}
\center{{
    \epsfig{figure=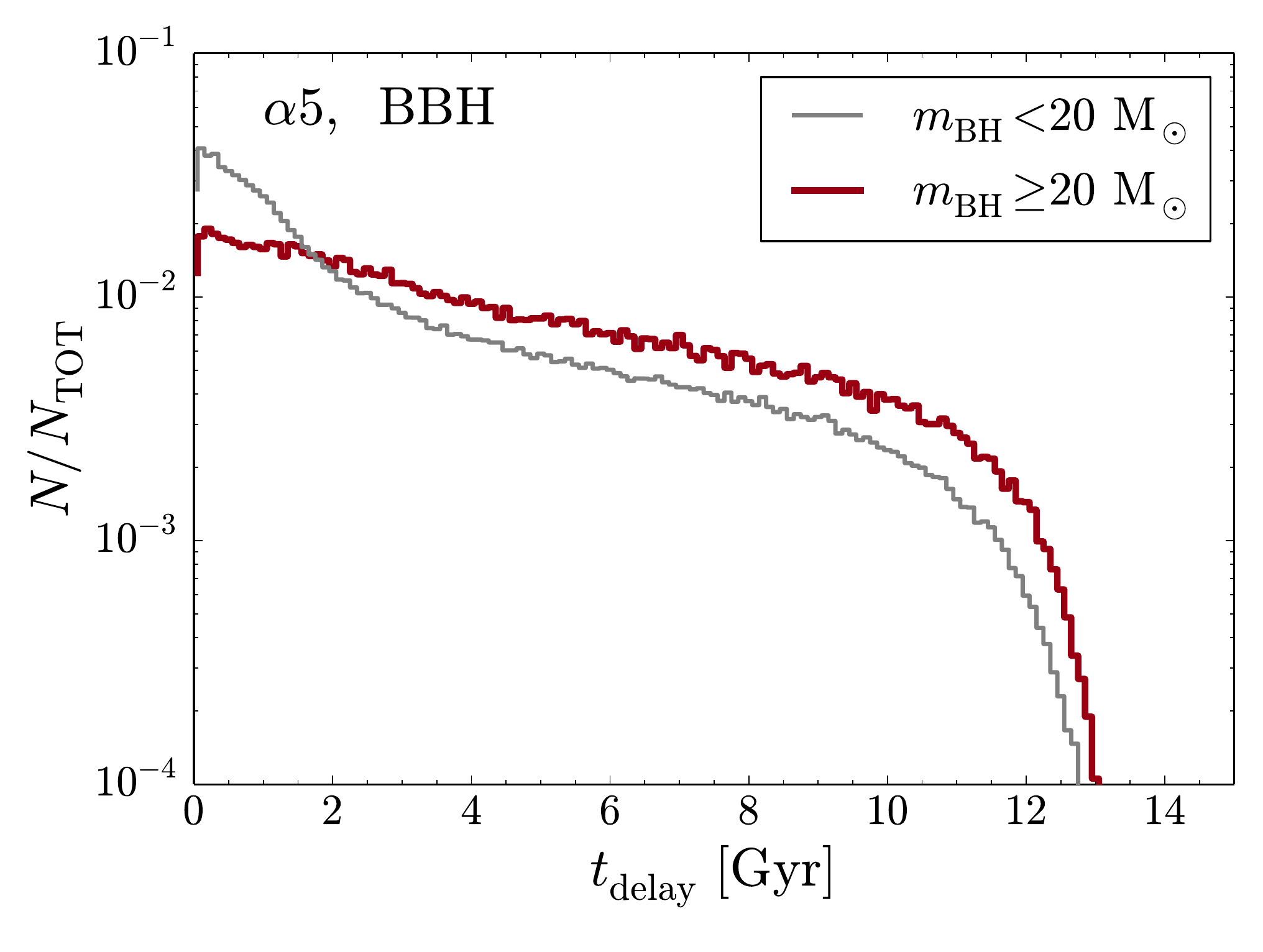,width=8cm} %was fig1.pdf
    \epsfig{figure=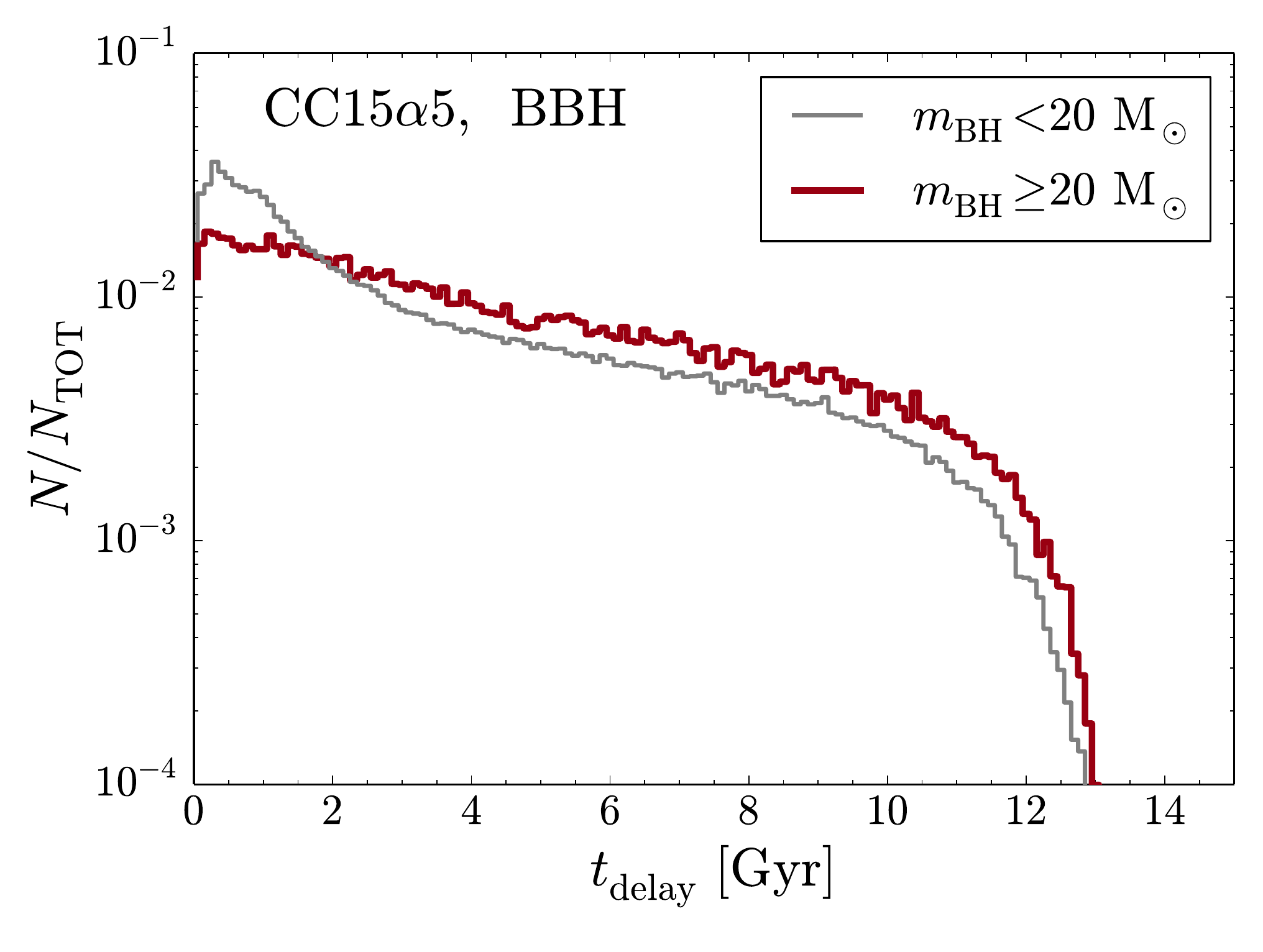,width=8cm} %was fig1.pdf
}}
\caption{Delay time $t_{\rm delay}$ (i.e. time elapsed from the formation of the stellar binary to the merger of the two compact objects)  of all merging BBHs (regardless of their merger redshift) in run $\alpha{}5$ (top) and  CC15$\alpha{}5$ (bottom). Thick dark red line: BBHs with mass $m_{\rm BH}\geq{}20$ M$_\odot$; thin grey line: BBHs with mass $m_{\rm BH}<20$ M$_\odot$. The number of BHs $N$ on the $y-$axis is normalized to the total number of BHs $N_{\rm TOT}$ in each histogram. \label{fig:BBHheavylight}
}
\end{figure}
%%%%%%%%%%%%%%%%%%%%%%%%%%%%%%%%%%%%%%%%%%%%%%%%%%%%%%%%%%%%%%%%%%%%%%%%%%%%%%%

%%%%%%%%%%%%%%%%%%%%%%%%%%%%%%%%%%% FIGURE 3 %%%%%%%%%%%%%%%%%%%%%%%%%%%%%%%%%%
\begin{figure}
\center{{
    \epsfig{figure=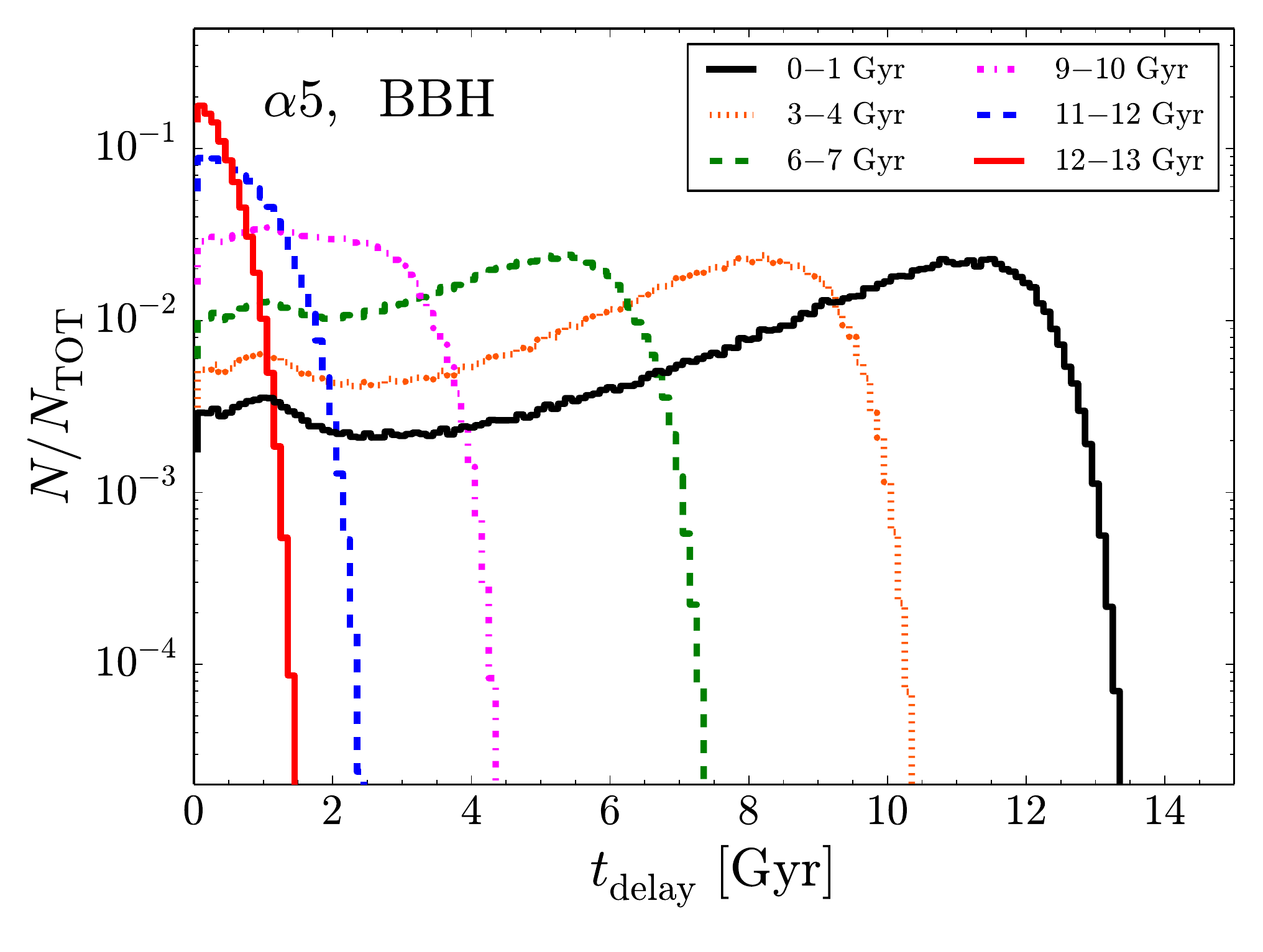,width=8cm} %was fig1.pdf
    \epsfig{figure=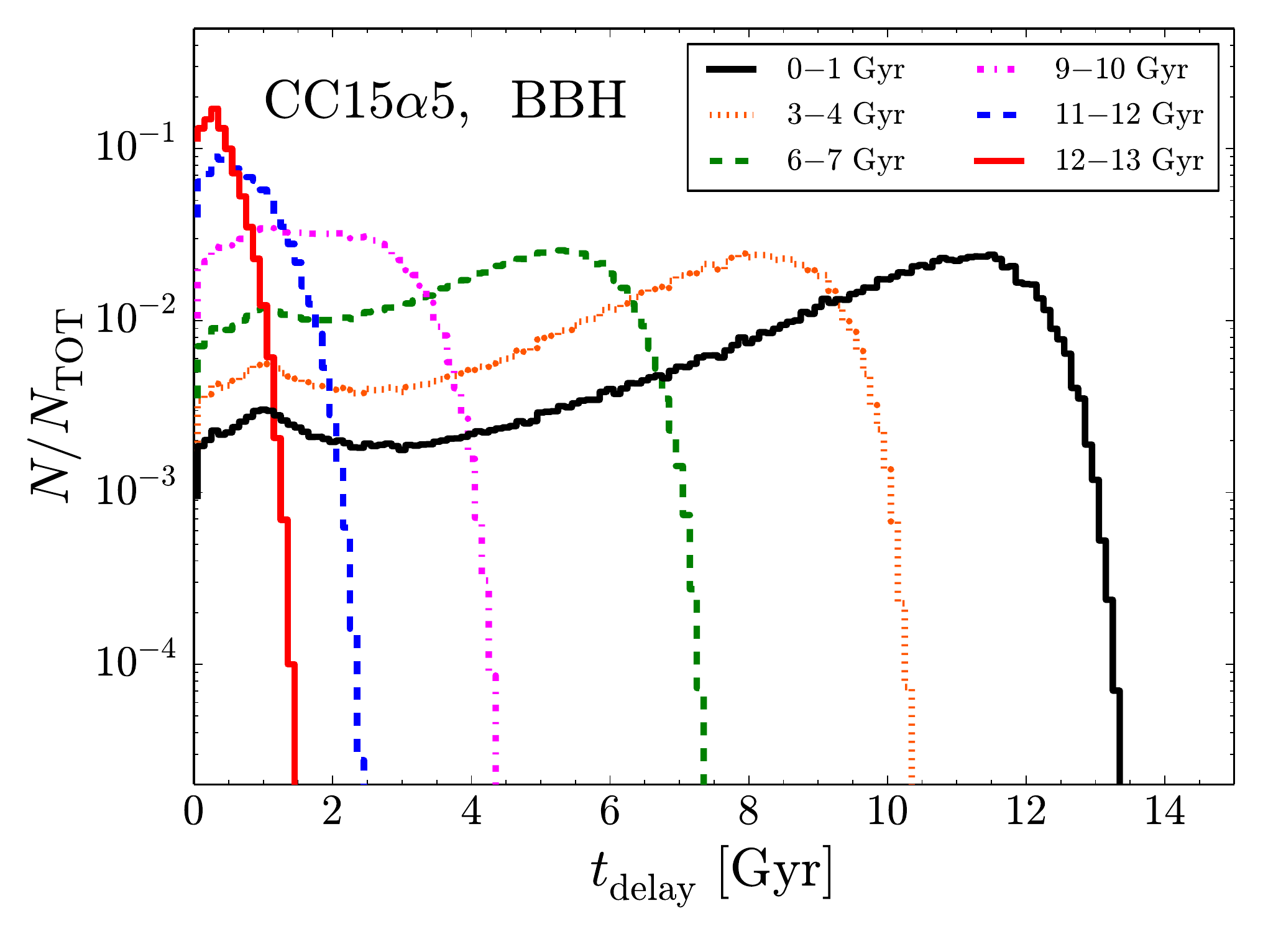,width=8cm} %was fig1.pdf
}}
\caption{Delay time  of BBHs in run $\alpha{}5$ (top) and  CC15$\alpha{}5$ (bottom) grouped by their merger time. Black solid line: BBHs merging $0-1$ Gyr ago ($z\leq{}0.08$); orange dotted line: BBHs merging $3-4$ Gyr ago ($z=0.26-0.37$); green dashed line: BBHs merging $6-7$ Gyr ago ($z=0.64-0.82$); magenta dot-dashed line:  BBHs merging $9-10$ Gyr ago ($z=1.35-1.78$); blue dashed line: BBHs merging $11-12$ Gyr ago ($z=2.43-3.65$); red solid line: BBHs merging $12-13$ Gyr ago ($z=3.65-7.15$).  The number of BHs $N$ on the $y-$axis is normalized to the total number of BHs $N_{\rm TOT}$ in each histogram. \label{fig:BBHdelay}
}
\end{figure}
%%%%%%%%%%%%%%%%%%%%%%%%%%%%%%%%%%%%%%%%%%%%%%%%%%%%%%%%%%%%%%%%%%%%%%%%%%%%%%%

%%%%%%%%%%%%%%%%%%%%%%%%%%%%%%%%%%% FIGURE 4 %%%%%%%%%%%%%%%%%%%%%%%%%%%%%%%%%%
\begin{figure}
\center{{
    \epsfig{figure=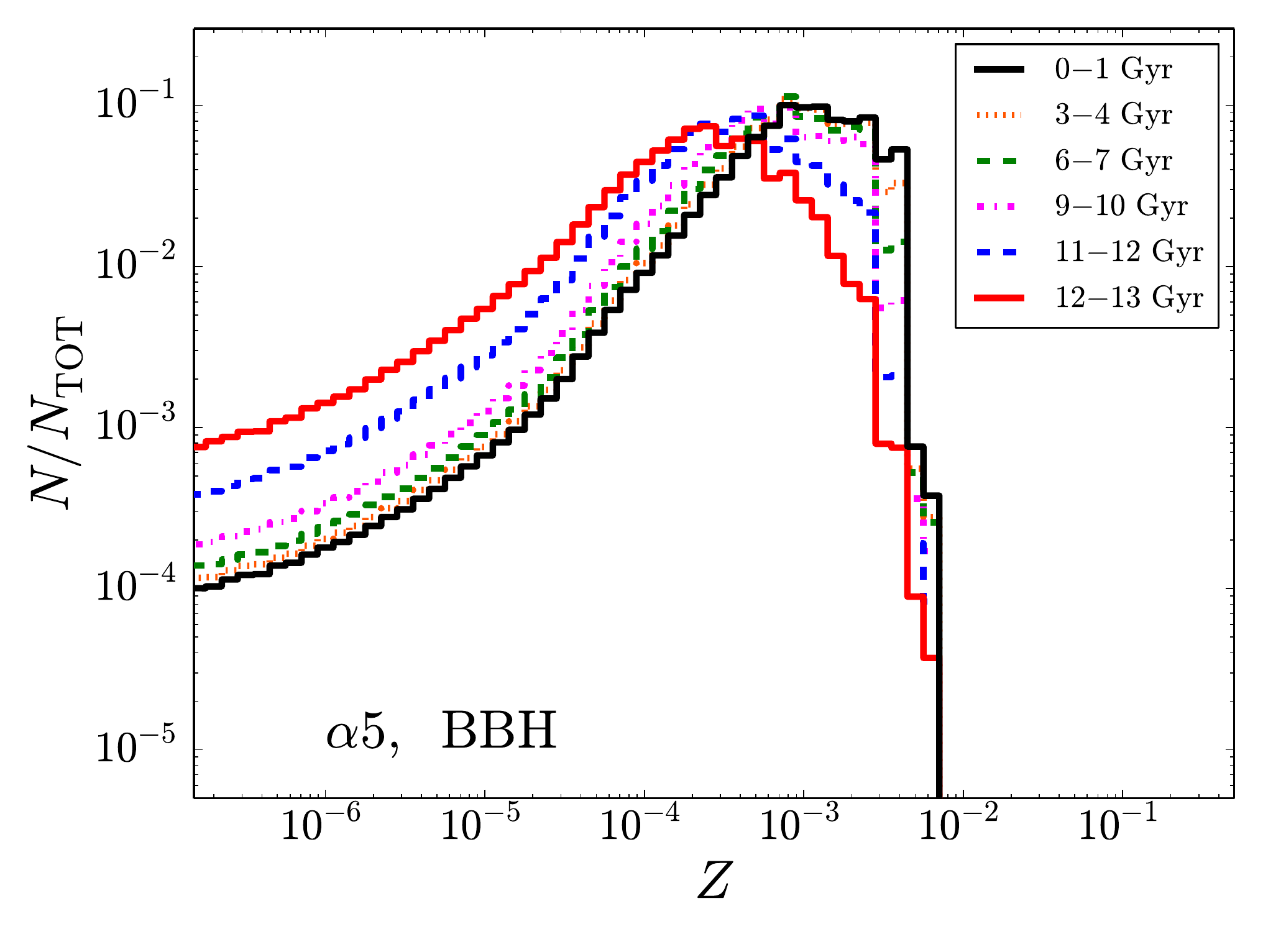,width=8cm} %was fig1.pdf
    \epsfig{figure=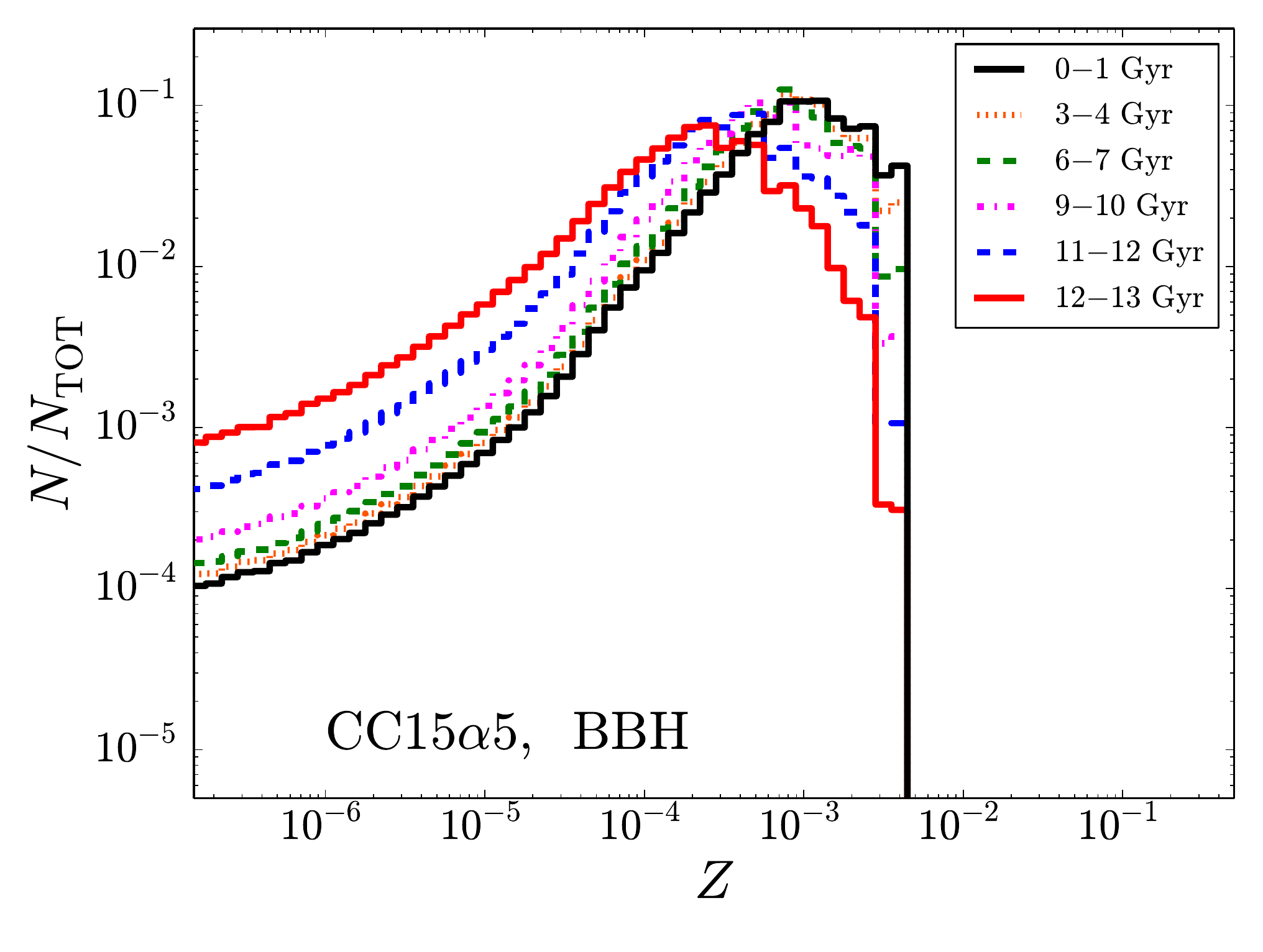,width=8cm} %was fig1.pdf
}}
\caption{Metallicity of BBH progenitors in run $\alpha{}5$ (top) and  CC15$\alpha{}5$ (bottom). Different lines are the same as in Figure~\ref{fig:BBHdelay}. The number of BHs $N$ on the $y-$axis is normalized to the total number of BHs $N_{\rm TOT}$ in each histogram.\label{fig:BBHZ}
}
\end{figure}
%%%%%%%%%%%%%%%%%%%%%%%%%%%%%%%%%%%%%%%%%%%%%%%%%%%%%%%%%%%%%%%%%%%%%%%%%%%%%%%

In this paper, we consider two different runs performed with \textsc{mobse} and already presented in \cite{giacobbo2018b}: run~$\alpha{}5$ and run~CC15$\alpha{}5$ (see Table~\ref{tab:table1}). In both runs, we assume CE efficiency $\alpha{}=5$. We have chosen $\alpha{}=5$ as a fiducial value, because Figure~15 of \cite{giacobbo2018b} shows that the BNS merger rate inferred from GW170817 \citep{abbottO2} is difficult to match  if we assume a lower value of $\alpha{}$.

  Runs $\alpha{}5$ and CC15$\alpha5$ differ only by the magnitude of the natal kick of core-collapse SNe: $\sigma{}_{\rm CCSN}=265$ and $15$ km s$^{-1}$ in run $\alpha{}5$ and run~CC$15\alpha{}5$, respectively. In \cite{giacobbo2018b}, we have shown that the local merger rate of BNSs derived from run~CC15$\alpha{}5$ is consistent with the one inferred from GW170817, while the local BNS merger rate derived from run $\alpha{}5$ is only marginally consistent with GW170817. The local merger rate of NSBHs and BBHs is consistent with the numbers derived by the LIGO-Virgo collaborations in both run~$\alpha{}5$ and CC15$\alpha{}5$ (see Figure 15 of \citealt{giacobbo2018b}). In the Appendix we discuss three supplementary runs where we consider different values of $\alpha{}$ ($=1,$ 3) and different assumptions for the natal kick.

For each run,  we have simulated  12 sub-sets of binaries with metallicity $Z=0.0002,$ 0.0004, 0.0008, 0.0012, 0.0016, 0.002, 0.004, 0.006, 0.008, 0.012, 0.016, and 0.02. In each sub-set  we  have simulated $10^7$ stellar binaries. Thus, each of the two runs is composed of $1.2\times{}10^8$ massive binaries.

For each binary, the mass of the primary is randomly drawn from a Kroupa initial mass function \citep{kroupa2001} ranging from 5 to 150 M$_\odot$, and the mass of the secondary is sampled according to the distribution $\mathcal{F}(q)\propto{}q^{-0.1}$ (where $q$ is the ratio between mass of the secondary and mass of the primary) in a range  $[0.1\,{}-1]\,{}m_{\rm p}$. The orbital period $P$ and the eccentricity $e$ are randomly extracted from the distribution $\mathcal{F}(P)\propto{}(\log_{10}{P})^{-0.55}$, with $0.15\leq{}\log_{10}{(P/{\rm day})}\leq{}5.5$, and $\mathcal{F}(e)\propto{}e^{-0.4}$, with $0\leq{}e<1$ \citep{sana2012}.

\subsection{Coupling with the Illustris cosmological simulation}\label{sec:section2.2}
The {\sc{Illustris-1}} is the highest resolution hydrodynamical simulation run in the frame of the Illustris project \citep{vogelsberger2014a,vogelsberger2014b,nelson2015}. In the following, we refer to it simply as the Illustris. It covers a comoving volume of $(106.5\,{}{\rm Mpc})^3$, and has an initial dark matter and baryonic matter mass resolution of $6.26\times{}10^6$ and $1.26\times{}10^6$ M$_\odot$, respectively \citep{vogelsberger2014a,vogelsberger2014b}. The Illustris includes a treatment for sub-grid physics (cooling, star formation, SNe, super-massive BH formation, accretion and merger, AGN feedback, etc), as described in \cite{vogelsberger2013}.

As for the cosmology, the Illustris adopts WMAP-9 results for the cosmological parameters \citep{hinshaw2013}, that is $\Omega{}_M = 0.2726$, $\Omega{}_\Lambda = 0.7274$, $\Omega{}_b = 0.0456$, and $H_0 = 100\,{}h$ km s$^{-1}$ Mpc$^{-1}$, with $h = 0.704$.

We combine the catalogues of merging BNSs, NSBHs and BBHs with the Illustris simulations through a Monte Carlo procedure, as first described in \cite{mapelli2017}. In particular, we extract all new-born star particles from the Illustris and we associate to each of these star particles a number $n_{\rm CO,\,{}i}$ of merging compact-object binaries (where the index $i=$ BBH, NSBH or BNS) based on the initial mass $M_{\rm Ill}$ and on the metallicity $Z_{\rm Ill}$ of each Illustris star particle. In particular, 
\begin{equation}
n_{\rm CO,\,{}i}=N_{\rm BSE,\,{}i}\,{}\frac{M_{\rm Ill}}{M_{\rm BSE}}\,{}f_{\rm corr}\,{}f_{\rm bin},
\end{equation}
where $M_{\rm BSE}$ is the total initial stellar mass of the population simulated with {\sc{mobse}} whose metallicity is closer to  $Z_{\rm Ill}$, while $N_{\rm BSE,\,{}i}$ is the number of merging compact-object binaries which form in the stellar population of mass $M_{\rm BSE}$. The term $f_{\rm corr}= 0.285$ is a correction factor, accounting for the fact that we actually simulate only primaries with zero-age main sequence mass $m_{\rm ZAMS}\ge{}5$ M$_\odot$, neglecting lower mass stars. The term $f_{\rm bin}$ accounts for the fact that we simulate only binary systems. Here we assume that 50 \% of stars are in binaries, thus $f_{\rm bin}=0.5$. We note that this is a simplifying assumption, because $f_{\rm bin}$ should also depend on stellar mass \citep{oshaughnessy2010}.  This dependence might be important when calculating the merger rate density, while it does not affect significantly the properties of merging binaries.

Finally, we randomly draw $n_{\rm CO,\,{}i}$ merging compact objects from the simulated population with mass $M_{\rm BSE}$ and we calculate the look-back time of the merger of each compact object binary  as
\begin{equation}\label{eq:eqmerge}
  t_{\rm merge}=t_{\rm form}-t_{\rm delay},
\end{equation}
  where $t_{\rm delay}$ is the delay time (i.e. the time elapsed between the formation of the stellar binary and the merger of the two compact objects) and $t_{\rm form}$ is the look back time at which the Illustris' particle has formed, calculated as
\begin{equation}
t_{\rm form}=\frac{1}{H_0}\int_0^{z_{\rm Ill}}\frac{1}{(1+z)\,{}\left[\Omega{}_M\,{}(1+z)^3+\Omega{}_\Lambda{}\right]^{1/2}}{\rm d}z,
\end{equation}
where the cosmological parameters are set to WMAP-9 values (for consistency with the Illustris) and $z_{\rm Ill}$ is the formation redshift of the Illustris' particle.  We have a minus sign in equation~\ref{eq:eqmerge} because both $t_{\rm merge}$ and $t_{\rm form}$ are look-back times.

\section{Results}\label{sec:results} 
\subsection{Binary black holes (BBHs)}
Figure~\ref{fig:BBHmass} shows the mass distribution of BBHs merging in the local Universe ($\leq{}1$ Gyr ago) and at higher redshift. In particular, we consider BBHs merging $\leq{}1$, $3-4$, $6-7$, $9-10$, $11-12$ and $12-13$ Gyr ago (corresponding to redshift $z\leq{}0.08$, $0.26-0.37$, $0.64-0.82$, $1.35-1.78$, $2.43-3.65$ and $3.65-7.15$, respectively).
From this Figure, it is apparent that the mass of BHs in merging BBHs evolves only mildly with redshift.

This might seem a bit surprising but comes from two facts. First, metal-rich binaries are much less efficient than metal-poor ones in producing merging BBHs \citep{belczynski2010b,giacobbo2018,klencki2018}. In particular, 
the number of mergers per unit stellar mass is approximately $3-4$ orders of magnitude higher at low metallicity than at solar metallicity, according to the population-synthesis runs adopted this paper  (see e.g. Figure~14 of \citealt{giacobbo2018b}). Thus, the contribution of metal-rich stars to the merger rate is so small that most merging BBHs come from metal-poor stars even in the local Universe.
Second, metallicity in the Universe is quite patchy: even at low redshift we can find a number of galaxies with very low metallicity, while even at high redshift the most massive galaxies have already reached a significantly high metallicity (see e.g. \citealt{maiolinomannucci2018} for a recent review).

The main difference between BBHs merging in the local Universe and BBHs merging $>12$ Gyr ago concerns BBHs with mass $\sim{}20-35$ M$_\odot$. BHs with mass $>20$ M$_\odot$ are more common in BBHs that merge $\leq{}1$ Gyr ago than in BBHs merging $\geq{}12$ Gyr ago. This might be surprising because in our model more massive BHs are produced by metal-poor stars. On the other hand, this can be explained with a different delay time: the most massive BBHs tend to have longer delay time than light BBHs. This is apparent from Figure~\ref{fig:BBHheavylight}, where we show the delay time distribution of all simulated merging BBHs (regardless of their merger redshift) distinguishing between massive BHs ($m_{\rm BH}\geq{}20$ M$_\odot$) and light BHs ($m_{\rm BH}<20$ M$_\odot$): there is a clear dearth of massive BHs with $t_{\rm delay}\lesssim{}1$ Gyr. Thus, even if they form preferentially in the early Universe, massive BHs tend to merge locally with a long delay time.

 It is important to note that the distributions of delay times  shown in Figure~\ref{fig:BBHheavylight} refer to all BBHs merging within the cosmological simulation, regardless of when their progenitor stars formed: this is conceptually different from the delay time distribution of a coeval stellar population. The latter scales approximately as $t^{-1}$ \citep{oshaughnessy2010,dominik2012,belczynski2016}, with a moderate dependence on the assumed population-synthesis parameters (see e.g. Figure~6 of \citealt{giacobbo2018}).

Finally, Figure~\ref{fig:BBHmass} does not show any significant difference between the two  fiducial population-synthesis models we have considered ($\alpha{}5$ and CC15$\alpha{}5$).  In the Appendix, we discuss the impact on BBH masses of the CE parameter $\alpha{}$ and of the natal kick assumption. Our supplementary runs confirm that BBH mass changes very mildly with redshift. In particular, we find that if $\alpha{}=1$ even the slight dependence of BBH mass on redshift disappears.

%*BBH mass does not evolve significantly with redshift. a little bit surprising but comes from 2 things:
%-metal-rich stars produce a very small number of merging BHs thus most merging BHs come mostly from metal-poor stars even in local universe (long tdelay)
%-metallicity is patchy: even at low z metallicity is low in some galaxies (depends on illustris thus we must check with other metallicity evolution)

%* BBHH mass more massive BHs merge at 0-1 Gyr than 11-12, because massive BHs have longer tdelay

The distribution of delay times of BBHs grouped by different merger time (Figure~\ref{fig:BBHdelay}) confirms our analysis on BBH masses: most BBHs merging in the last Gyr form in the early Universe and have an extremely long delay time. Table~\ref{tab:table2} shows that $>97$\% of all BBHs merging in the last Gyr have a delay time $>1$ Gyr, and $\sim{}50$\% have a delay time $>10$ Gyr (i.e. they formed $>10$ Gyr ago). This result is in good agreement with previous work \citep{dominik2013,dominik2015,mapelli2017}.

The contribution of metal-rich binaries with short delay time  to the local BBH merger rate is very minor (see also \citealt{mapelli2017,mapelli2018,mapelli2018b}). Thus, the distribution of BH masses we can reconstruct from GW detections of BBHs merging in the local Universe reflects the distribution of BH masses formed several Gyr ago, rather than the local distribution of BH masses. This might be the key to explain the difference between the mass distribution of BHs in local X-ray binaries \citep{ozel2010,farr2011} and the mass distribution of BHs in GW events.

If we look at the metallicity of stellar progenitors (Figure~\ref{fig:BBHZ}), we find a significant shift between the peak metallicity of the progenitors of BBHs which merge $12-13$ Gyr ago ($Z_{\rm peak}\sim{}2-3\times{}10^{-4}$) and that of BBHs merging $\leq{}1$ Gyr ago ($Z_{\rm peak}\sim{}10^{-3}$). At all redshifts, merging BBHs with metallicity $\gtrsim{}4-6\times{}10^{-3}$ are extremely rare (see also \citealt{mapelli2017,mapelli2018}).

%* bulk of merging metallicity of course is lower for tmerg = 11-12 Gyr than for tmerg =0-1 Gyr but peak shift only little from $10^{-3}$ to $4-5\times{}10^{-4}$

Stars with metallicity $Z\ll{}10^{-5}$ (i.e. population~III stars) give a small contribution even to BBHs merging at high redshift, consistent with previous work \citep{hartwig2016,belczynski2017}. However, we warn that neither our population synthesis models include a specific treatment for population~III stars nor the Illustris simulation contains any specific sub-grid model to describe them, and thus we cannot specifically quantify their impact in the current study. 

%* metallicity $\ll{}10^{-5}$ gives negligible contribution

%* most BBHs merging at low Z (0-1 Gyr) have extremely long tdelay 

\subsection{Neutron star - black hole binaries (NSBHs)}
%%%%%%%%%%%%%%%%%%%%%%%%%%%%%%%%%%% FIGURE 5 %%%%%%%%%%%%%%%%%%%%%%%%%%%%%%%%%%
\begin{figure*}
\center{{
    \epsfig{figure=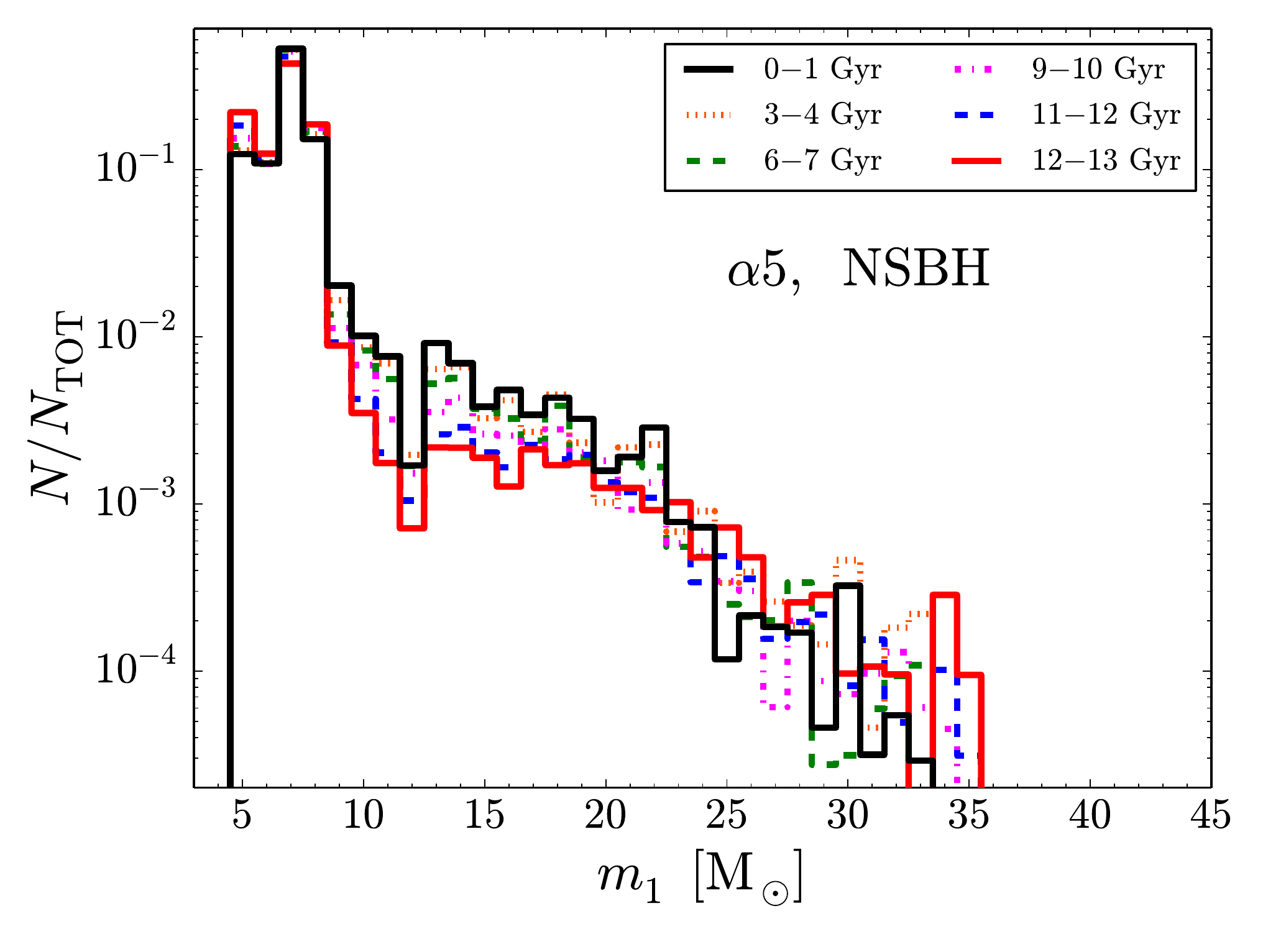,width=7cm} %was fig1.pdf
        \epsfig{figure=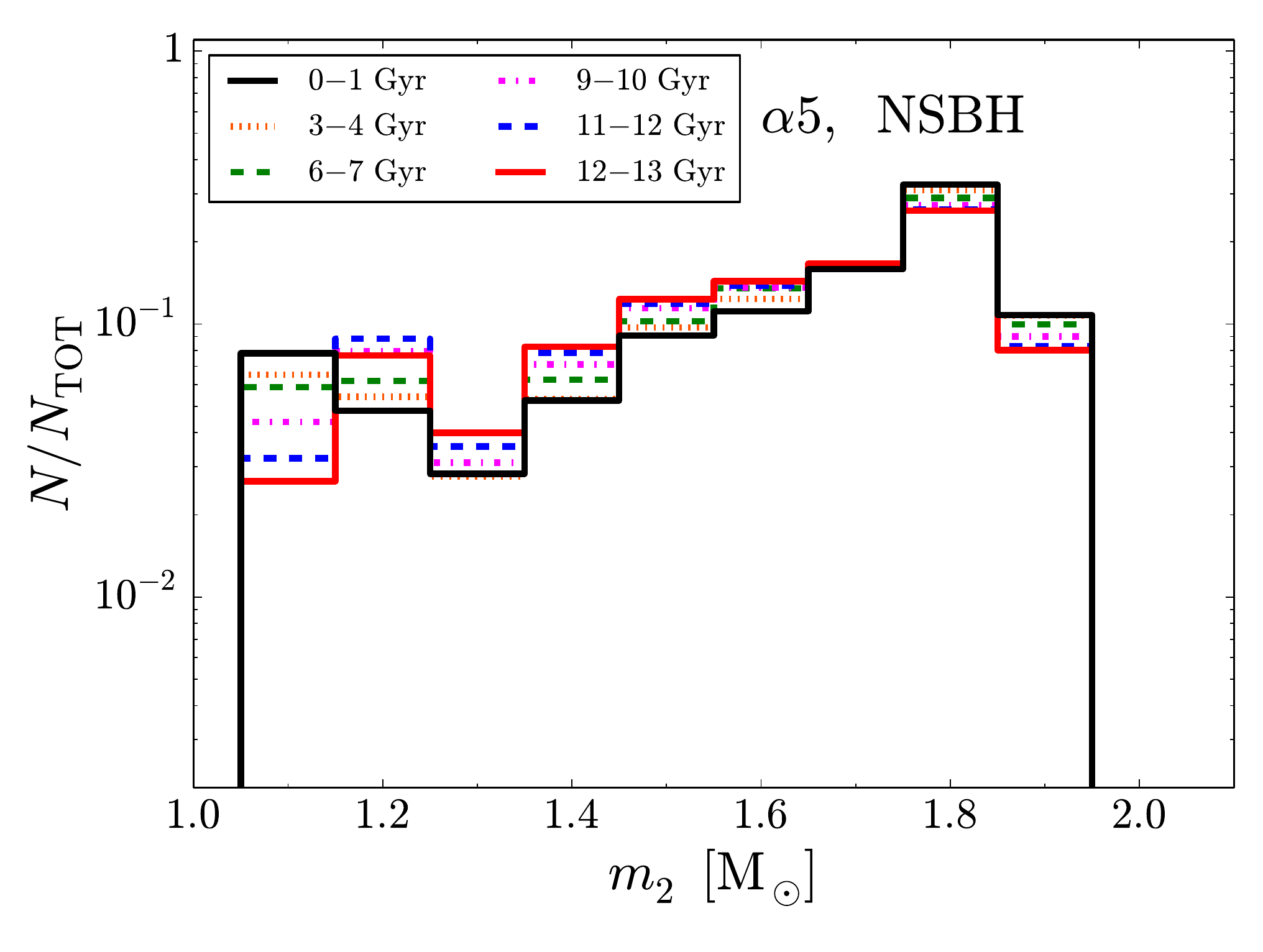,width=7cm} %was fig1.pdf
    \epsfig{figure=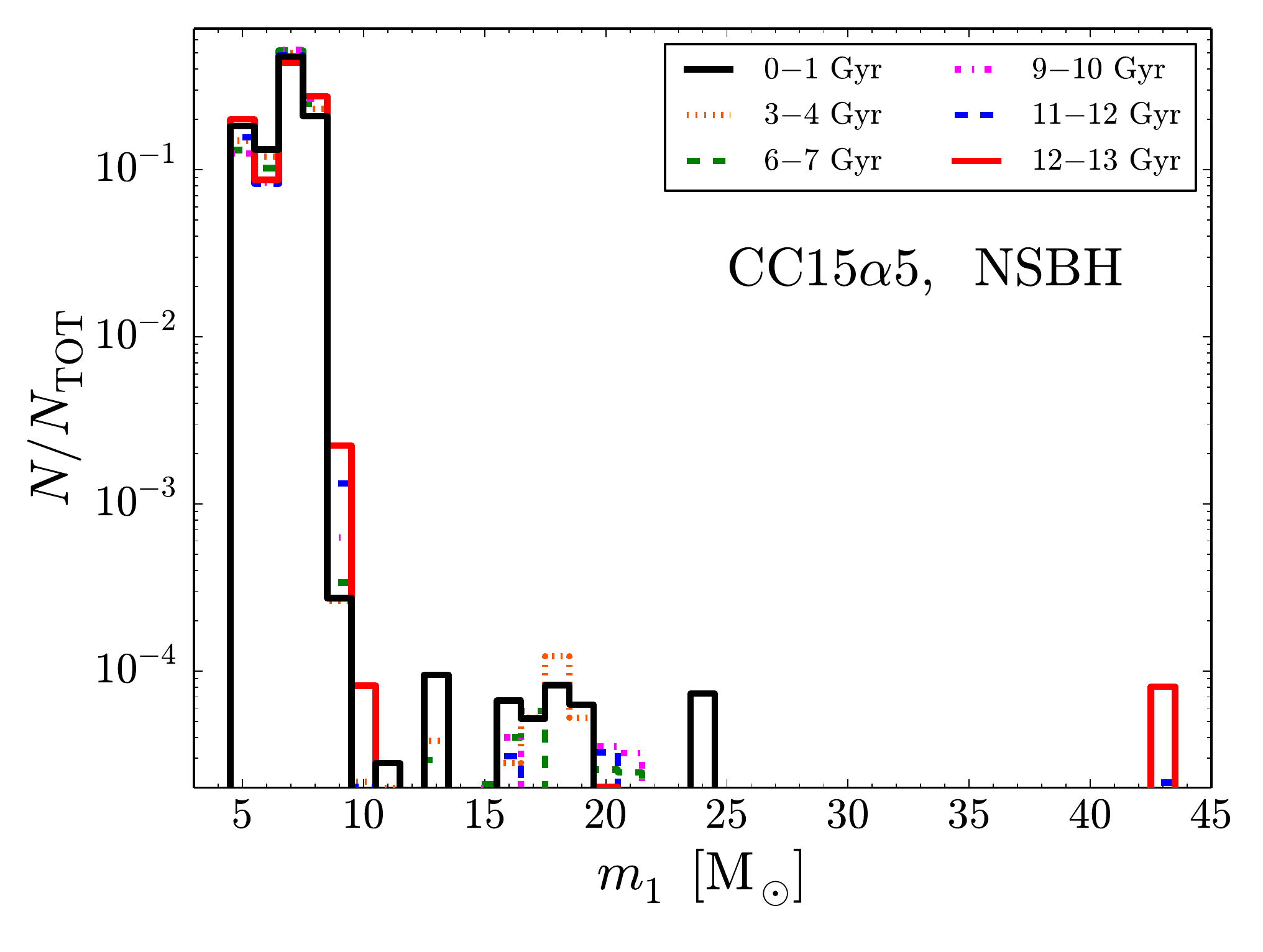,width=7cm} %was fig1.pdf
    \epsfig{figure=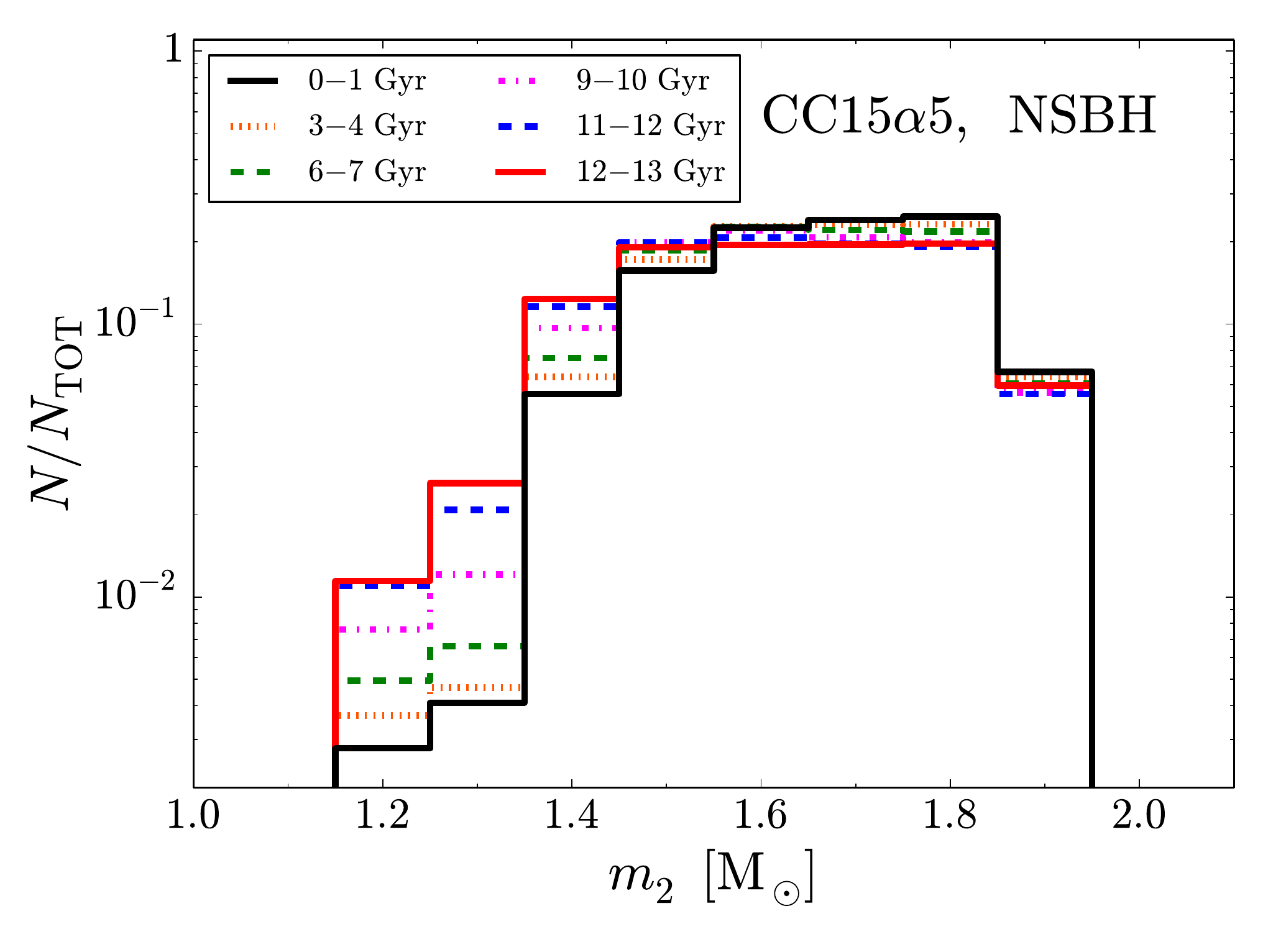,width=7cm} %was fig1.pdf
}}
\caption{Mass of the BH (left-hand panels) and of the NS (right-hand panel) in merging NSBHs. Black solid line: NSBHs merging $0-1$ Gyr ago ($z\leq{}0.08$); orange dotted line: NSBHs merging $3-4$ Gyr ago ($z=0.26-0.37$); green dashed line: NSBHs merging $6-7$ Gyr ago ($z=0.64-0.82$);  magenta dot-dashed line:  NSBHs merging $9-10$ Gyr ago ($z=1.35-1.78$); blue dashed line: NSBHs merging $11-12$ Gyr ago ($z=2.43-3.65$); red solid line: NSBHs merging $12-13$ Gyr ago ($z=3.65-7.15$). Top panels: run $\alpha{}5$. Bottom panels: run CC15$\alpha{}5$. The number of compact objects $N$ on the $y-$axis is normalized to the total number of compact objects $N_{\rm TOT}$ in each histogram.\label{fig:NSBHmass}
}
\end{figure*}
%%%%%%%%%%%%%%%%%%%%%%%%%%%%%%%%%%%%%%%%%%%%%%%%%%%%%%%%%%%%%%%%%%%%%%%%%%%%%%%

%%%%%%%%%%%%%%%%%%%%%%%%%%%%%%%%%%% FIGURE 6 %%%%%%%%%%%%%%%%%%%%%%%%%%%%%%%%%%
\begin{figure}
\center{{
    \epsfig{figure=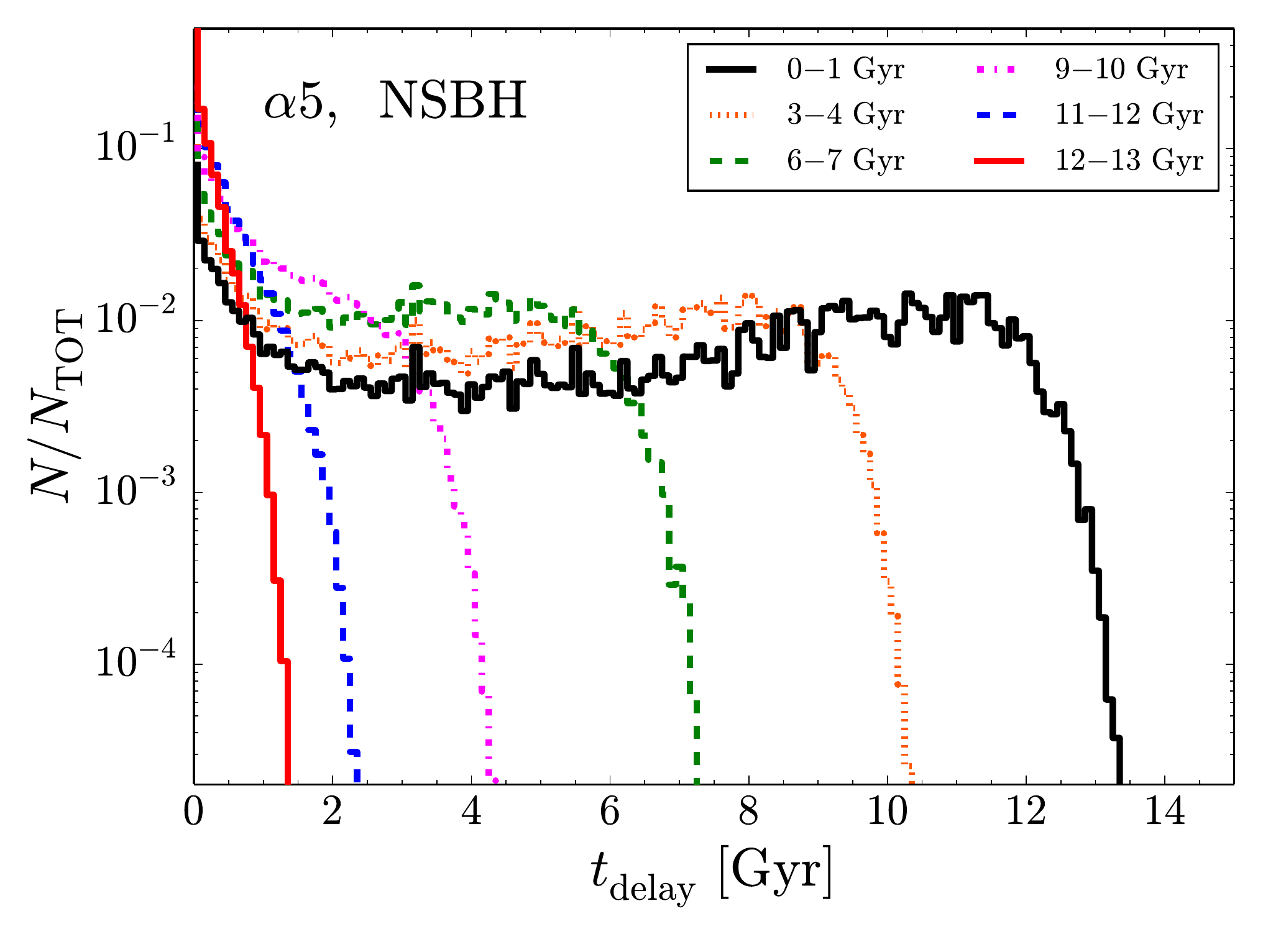,width=8cm} %was fig1.pdf
    \epsfig{figure=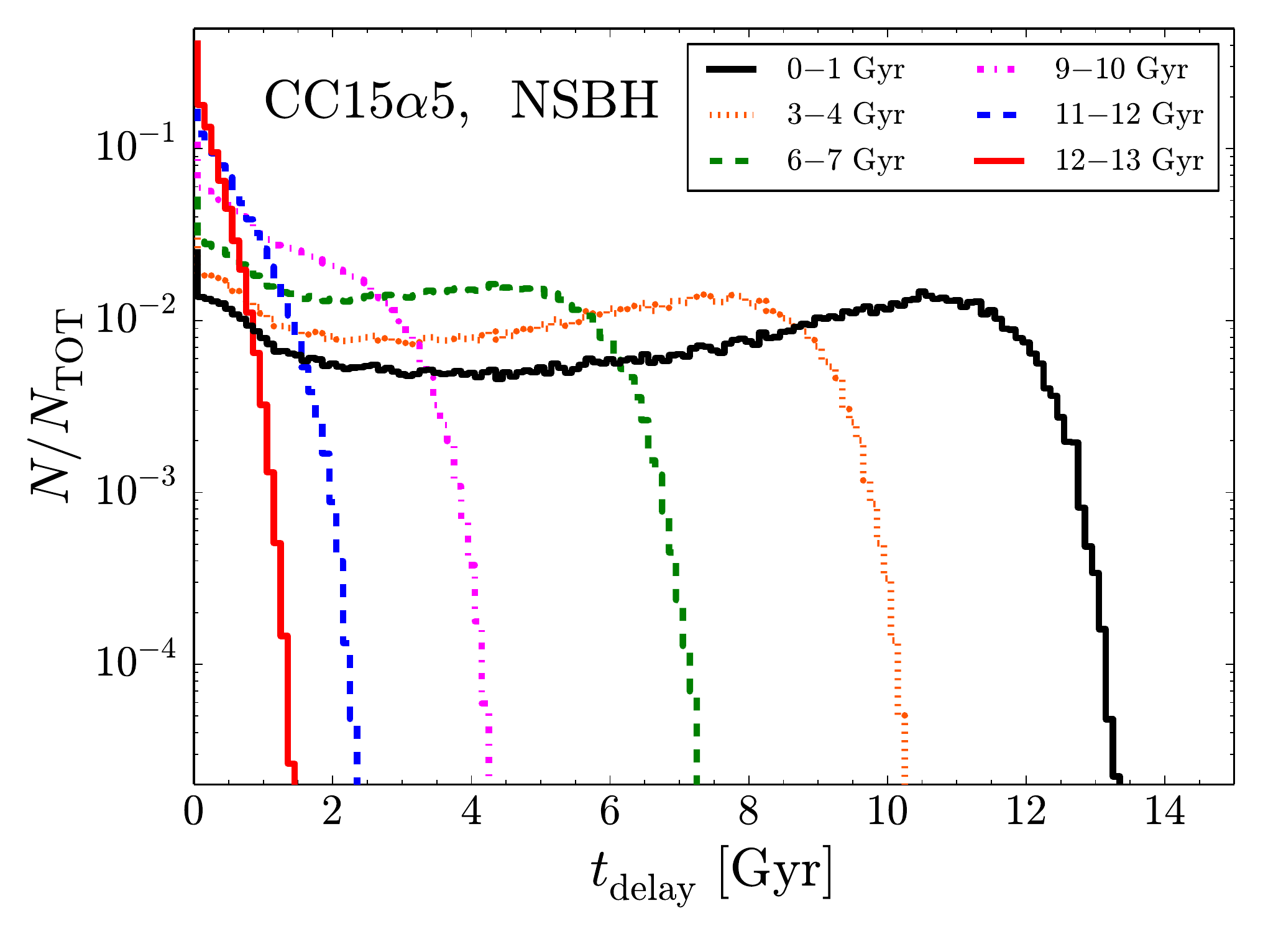,width=8cm} %was fig1.pdf
}}
\caption{Same as Fig.~\ref{fig:BBHdelay} but for NSBHs. \label{fig:NSBHdelay}
}
\end{figure}
%%%%%%%%%%%%%%%%%%%%%%%%%%%%%%%%%%%%%%%%%%%%%%%%%%%%%%%%%%%%%%%%%%%%%%%%%%%%%%%

%%%%%%%%%%%%%%%%%%%%%%%%%%%%%%%%%%% FIGURE 7 %%%%%%%%%%%%%%%%%%%%%%%%%%%%%%%%%%
\begin{figure}
\center{{
    \epsfig{figure=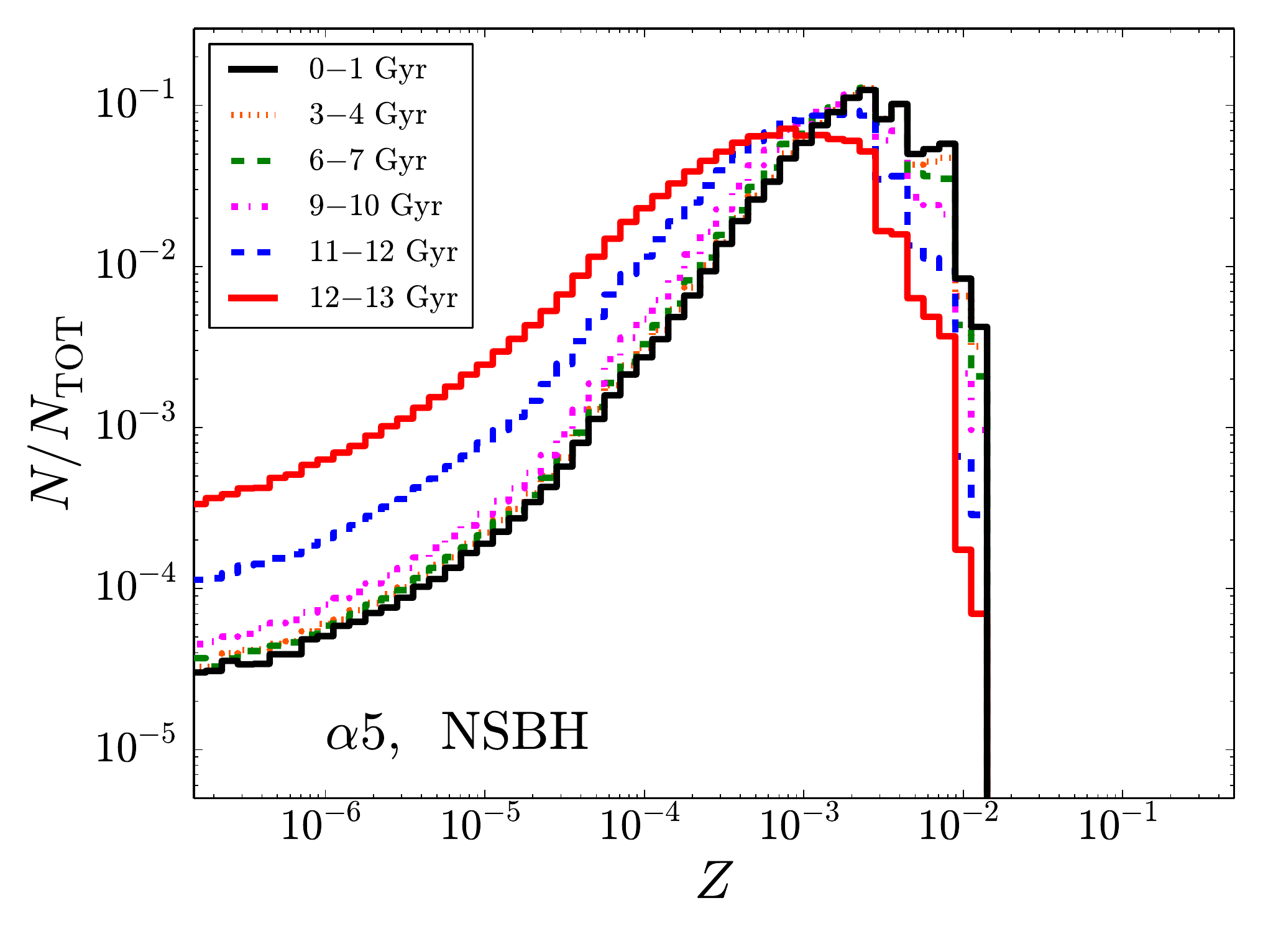,width=8cm} %was fig1.pdf
    \epsfig{figure=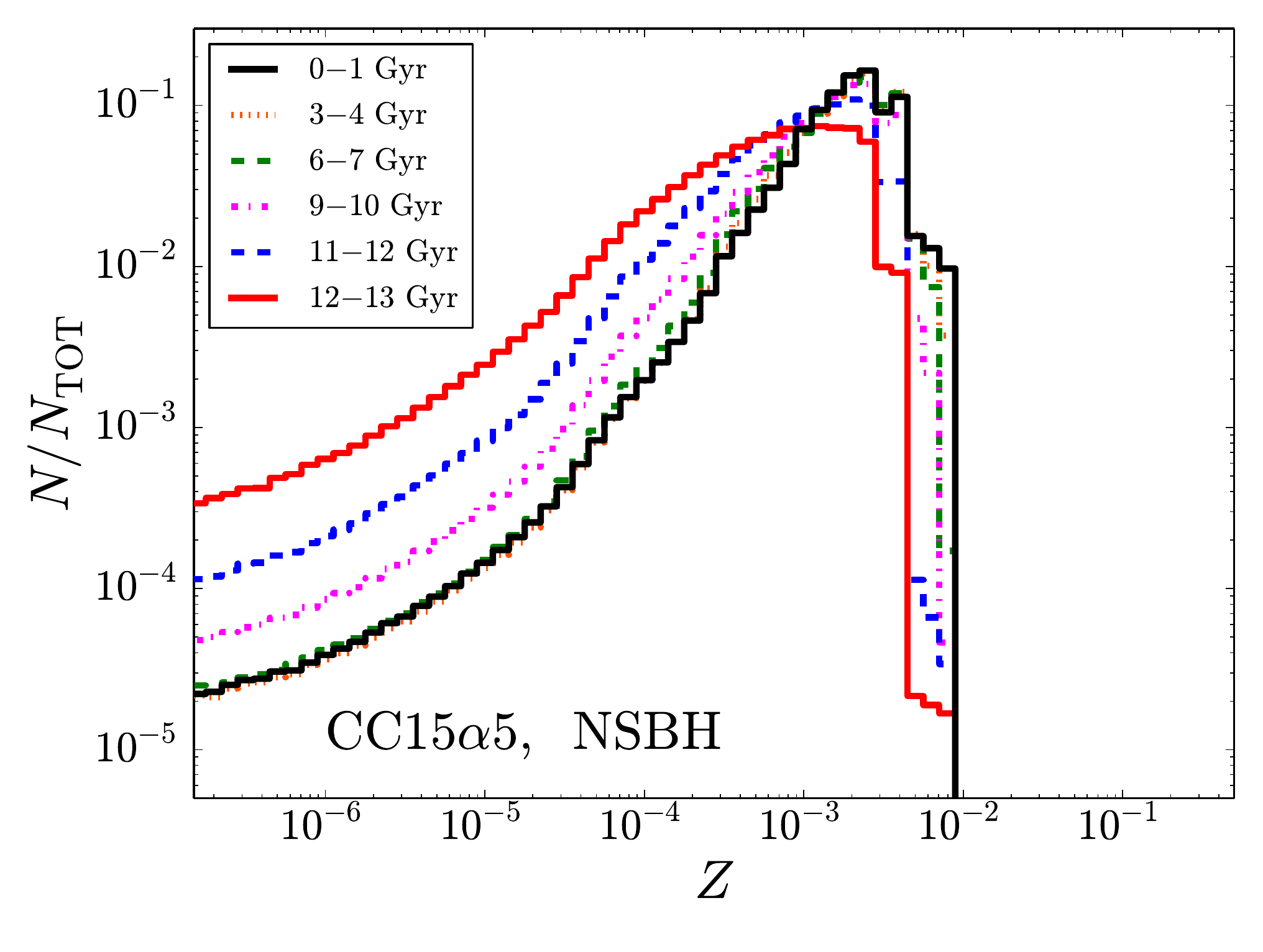,width=8cm} %was fig1.pdf
}}
\caption{Same as Fig.~\ref{fig:BBHZ} but for NSBHs.\label{fig:NSBHZ}
}
\end{figure}
%%%%%%%%%%%%%%%%%%%%%%%%%%%%%%%%%%%%%%%%%%%%%%%%%%%%%%%%%%%%%%%%%%%%%%%%%%%%%%%

%* also mass of NSBH does not evolve much with z for same reason
Even the mass of BHs in merging NSBHs does not change significantly with time (see the left-hand panel of Figure~\ref{fig:NSBHmass}). Again, this originates from the fact that the merger efficiency of NSBHs born from metal-poor progenitors is orders of magnitude higher than the merger efficiency of NSBHs born from metal-rich progenitors (see \citealt{giacobbo2018b}).

%*low mass BHs are more frequent than high mass BHs in NSBHs. this is dramatically true for cc15alpha5 (because low kicks do not unbind even the lighter binaries) but still holds for alpha5

%*high mass NS are more frequent than low mass NS in NSBHs, regardless of kick. Accretion tends to balance binary mass? Ask nicola.

Light BHs are more frequent than massive BHs in NSBHs: most BHs in merging NSBHs have mass $\leq{}10$ M$_\odot$, regardless of the merger time. The dearth of massive BHs is particularly strong in the case of run CC$15\alpha{}5$. %(because low kicks do not unbind even the lighter binaries), but still holds for $\alpha5$.

From the right-hand panel of Figure~\ref{fig:NSBHmass}, we see that massive neutron stars (NSs, $m_{\rm NS}>1.4$ M$_\odot$) are more common than low-mass NSs in NSBHs. This holds in both runs, although the dearth of light NSs is even stronger in run CC15$\alpha{}5$.

Thus, merging NSBHs tend to have the maximum possible mass ratio between the NS and BH ($m_{\rm NS}/m_{\rm BH}\sim{}0.1-0.4$ in our models which enforce the mass gap between NSs and BHs). This can be explained with the fact that non-conservative mass transfer and CE in close binaries tend to lead to compact objects of similar mass.
%Accretion tends to balance binary mass?

%in NSBH has same trend than for BBHs but with even less difference between NSBHs merging at high and low redshift + the bulk is at larger $Z\sim{}2-3\times{}10^{-3}$ - no significant difference between ccalpha and alpha

%*tdelay: much steeper tdelay than BBH - means that although NSBHs with long tdelay merge today there is also  a contribution from systems with short tdelay effect of higher metallicity

The distribution of NSBH delay times (Figure~\ref{fig:NSBHdelay}) is much steeper than that of BBHs: there is a significantly larger number of merging NSBHs with short delay times with respect to  BBHs \citep{dominik2013,dominik2015,mapelli2018b}. Table~\ref{tab:table2} shows that $\sim{}25-28$\% of NSBHs merging in the last Gyr formed $\geq{}10$ Gyr ago, but $\sim{}13-22$\% formed $<1$ Gyr ago: even NSBHs which formed in the last Gyr give an important contribution to the population of NSBHs merging in the local Universe.   %A large fraction of NSBHs merging  in the last Gyr formed $\sim{}10-12$ Gyr ago, but also systems which form in the last Gyr give an important contribution to the population. 

The metallicity distribution of progenitors of NSBHs follows the same trend as that of BBHs: the progenitors of NSBHs merging $\geq{}12$ Gyr ago tend to be more metal-poor than those of NSBHs merging in the last Gyr. However, the offset between the two populations is less evident than in the case of BBHs (the peak metallicity being $Z_{\rm peak}\sim{}10^{-3}$ and $Z_{\rm peak}\sim{}1-3\times{}10^{-3}$ for NSBHs merging $12-13$ Gyr ago and NSBHs merging in the last Gyr, respectively). The progenitors of merging NSBHs have $Z\leq{}1-2\times{}10^{-2}$. There is no significant difference between run $\alpha{}5$ and CC15$\alpha{}5$.

 %%%%%%%%%%%%%%%%%%%%%%%%%%%%%%%%%% TABLE 2%%%%%%%%%%%%%%%%%%%%%%%%%%%%%%%%%%%%%
\begin{table}
\begin{center}
\caption{\label{tab:table2}
Percentage of systems that merge in the last Gyr and have $t_{\rm delay}>1$ Gyr or $>10$ Gyr.}
 \leavevmode
\begin{tabular}[!h]{lllll}
\hline
Run    & Binary type & $t_{\rm delay}>1$ Gyr & $t_{\rm delay}>10$ Gyr\\
\hline
$\alpha5$     & BBH & 0.97 & 0.49\\
CC15$\alpha5$ & BBH & 0.98 & 0.52\\
$\alpha5$     & NSBH & 0.78 & 0.25\\
CC15$\alpha5$ & NSBH & 0.87 & 0.28\\
$\alpha5$     & BNS & 0.17 & 0.02\\
CC15$\alpha5$ & BNS & 0.44 & 0.03\\
\hline
\end{tabular}
\begin{flushleft}
\footnotesize{Column 1: model name; column 2: type of merging binary (BBH, NSBH or BNS); column 3: percentage of binaries that merge in the last Gyr and have $t_{\rm delay}>1$ Gyr; column 4: percentage of binaries that merge in the last Gyr and have $t_{\rm delay}>10$ Gyr.}
\end{flushleft}
\end{center}
\end{table}
%%%%%%%%%%%%%%%%%%%%%%%%%%%%%%%%%%%%%%%%%%%%%%%%%%%%%%%%%%%%%%%%%%%%%%%%%%%%%%%%

%BHBH a5   fast (<1Gyr) 0.03, slow 0.97,  very slow (>10 Gyr) 0.49
%BHBH cca5 fast (<1Gyr) 0.02, slow 0.98,  very slow (>10 Gyr) 0.52
%NSBH a5   fast (<1Gyr) 0.22, slow 0.78,  very slow (>10 Gyr) 0.25
%NSBH cca5 fast (<1Gyr) 0.13, slow 0.87,  very slow (>10 Gyr) 0.28
%NSNS a5   fast (<1Gyr) 0.83, slow 0.17,  very slow (>10 Gyr) 0.02
%NSNS cca5 fast (<1Gyr) 0.56, slow 0.44,  very slow (>10 Gyr) 0.03

\subsection{Binary neutron stars (BNSs)}
%%%%%%%%%%%%%%%%%%%%%%%%%%%%%%%%%%% FIGURE 8 %%%%%%%%%%%%%%%%%%%%%%%%%%%%%%%%%%
\begin{figure*}
\center{{
    \epsfig{figure=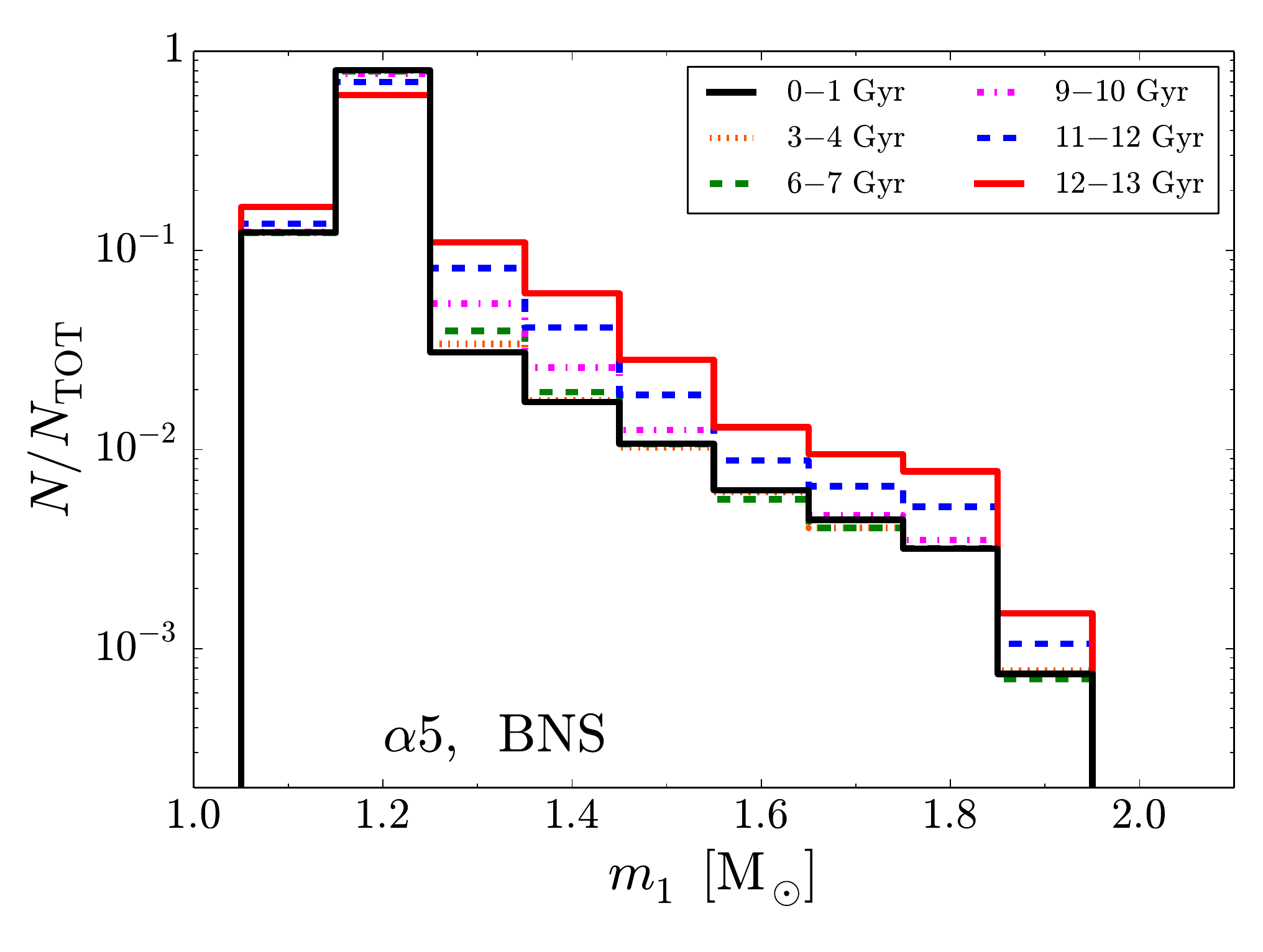,width=7cm} %was fig1.pdf
        \epsfig{figure=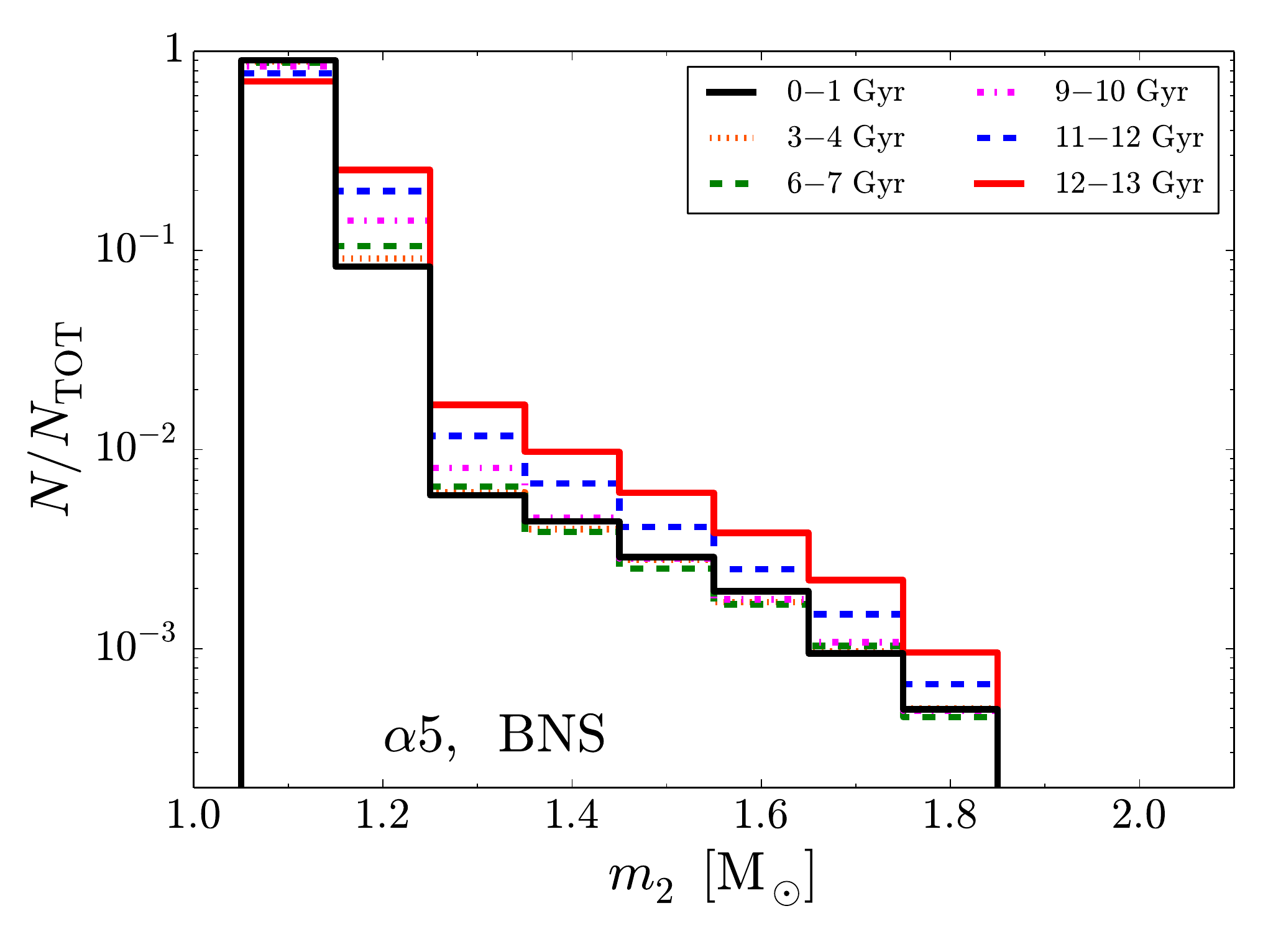,width=7cm} %was fig1.pdf
    \epsfig{figure=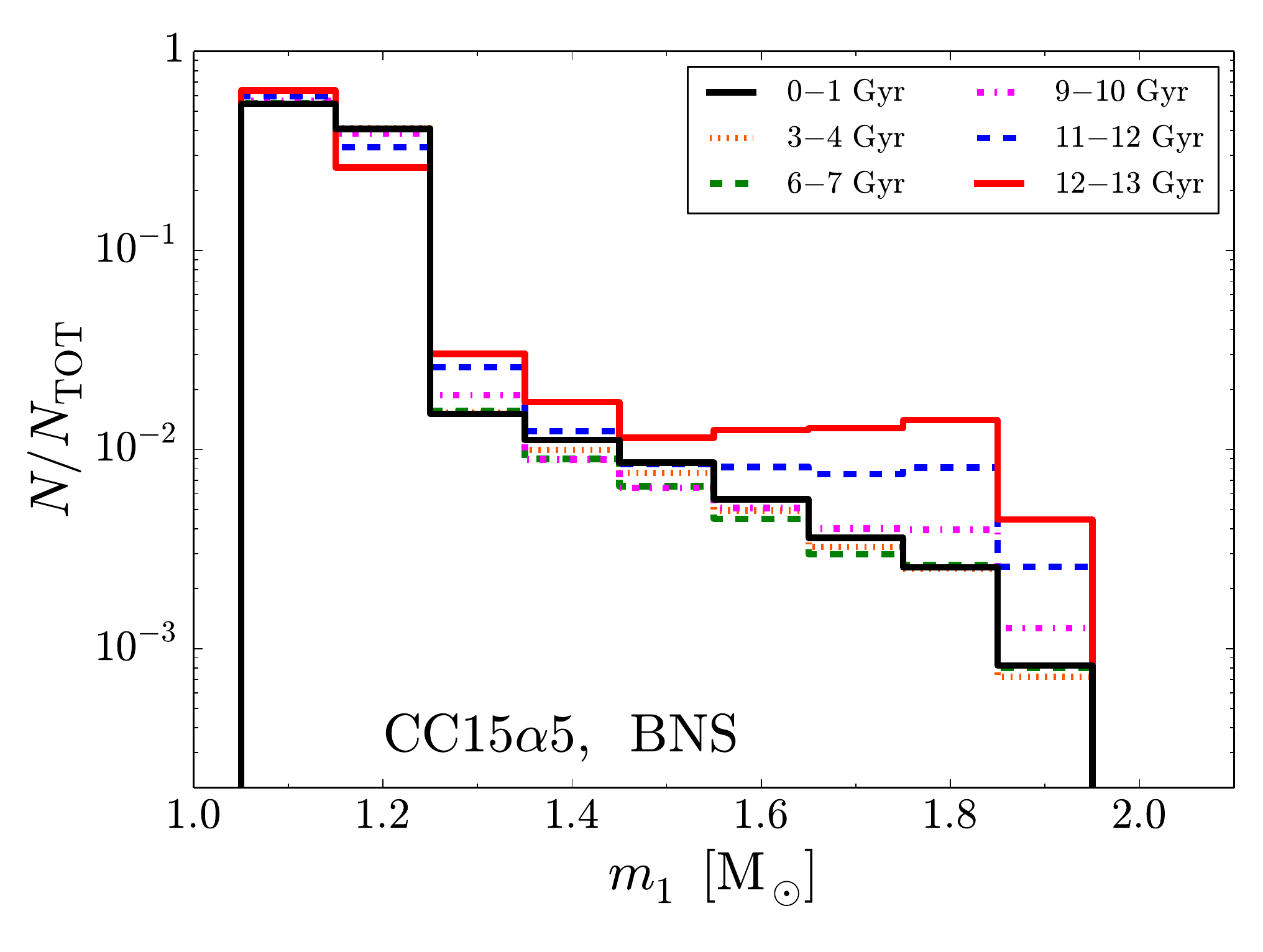,width=7cm} %was fig1.pdf
    \epsfig{figure=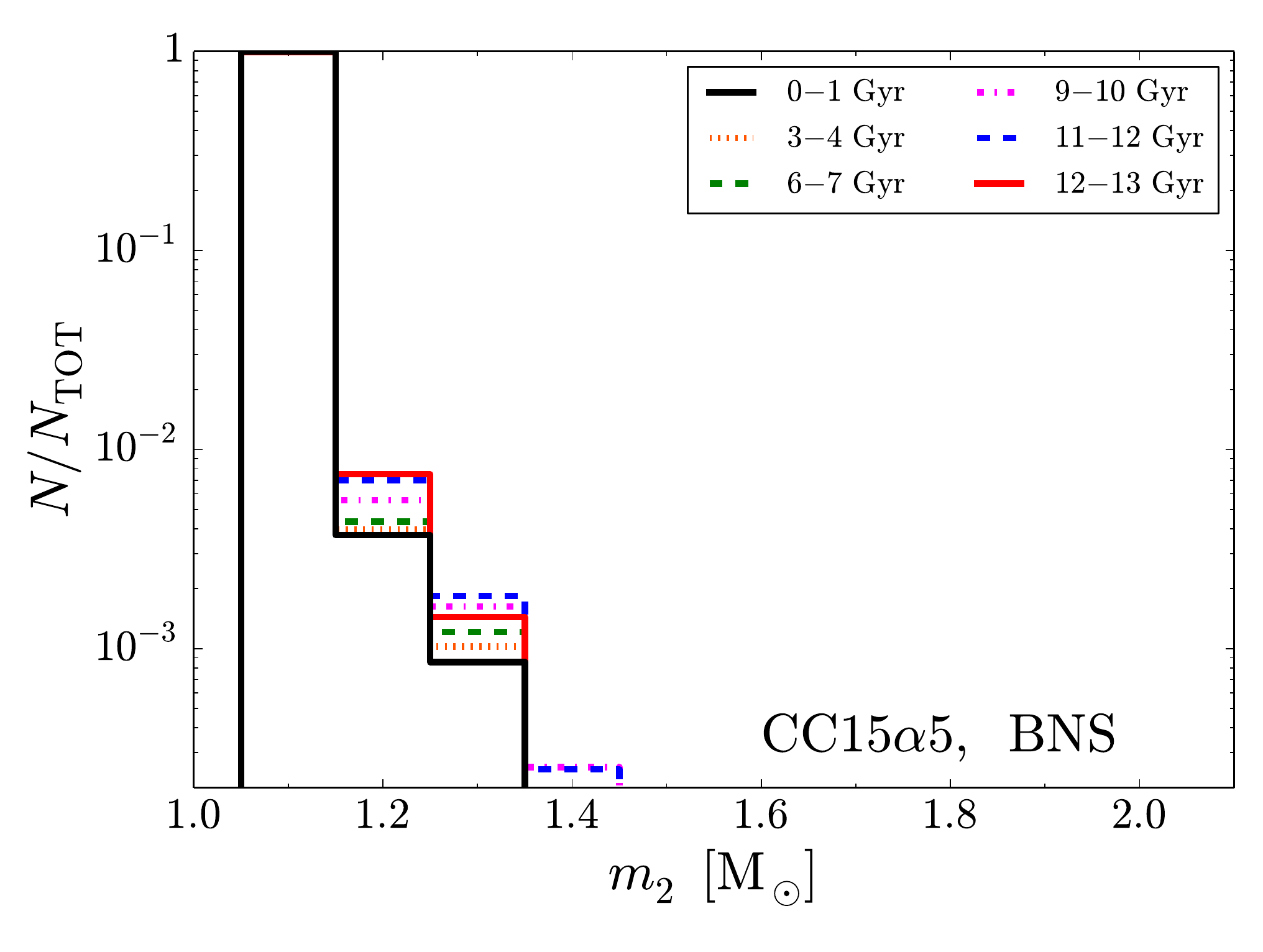,width=7cm} %was fig1.pdf
}}
\caption{Mass of the primary NS (left-hand panels) and of the secondary NS (right-hand panel) in merging BNSs. Black solid line: BNSs merging $0-1$ Gyr ago ($z\leq{}0.08$); orange dotted line: BNSs merging $3-4$ Gyr ago ($z=0.26-0.37$); green dashed line: BNSs merging $6-7$ Gyr ago ($z=0.64-0.82$);  magenta dot-dashed line:  BNSs merging $9-10$ Gyr ago ($z=1.35-1.78$); blue dashed line: BNSs merging $11-12$ Gyr ago ($z=2.43-3.65$); red solid line: BNSs merging $12-13$ Gyr ago ($z=3.65-7.15$). Top panels: run $\alpha{}5$. Bottom panels: run CC15$\alpha{}5$. The number of NSs $N$ on the $y-$axis is normalized to the total number of NSs $N_{\rm TOT}$ in each histogram.\label{fig:BNSmass}
}
\end{figure*}
%%%%%%%%%%%%%%%%%%%%%%%%%%%%%%%%%%%%%%%%%%%%%%%%%%%%%%%%%%%%%%%%%%%%%%%%%%%%%%%

%%%%%%%%%%%%%%%%%%%%%%%%%%%%%%%%%%% FIGURE 9 %%%%%%%%%%%%%%%%%%%%%%%%%%%%%%%%%%
\begin{figure}
\center{{
    \epsfig{figure=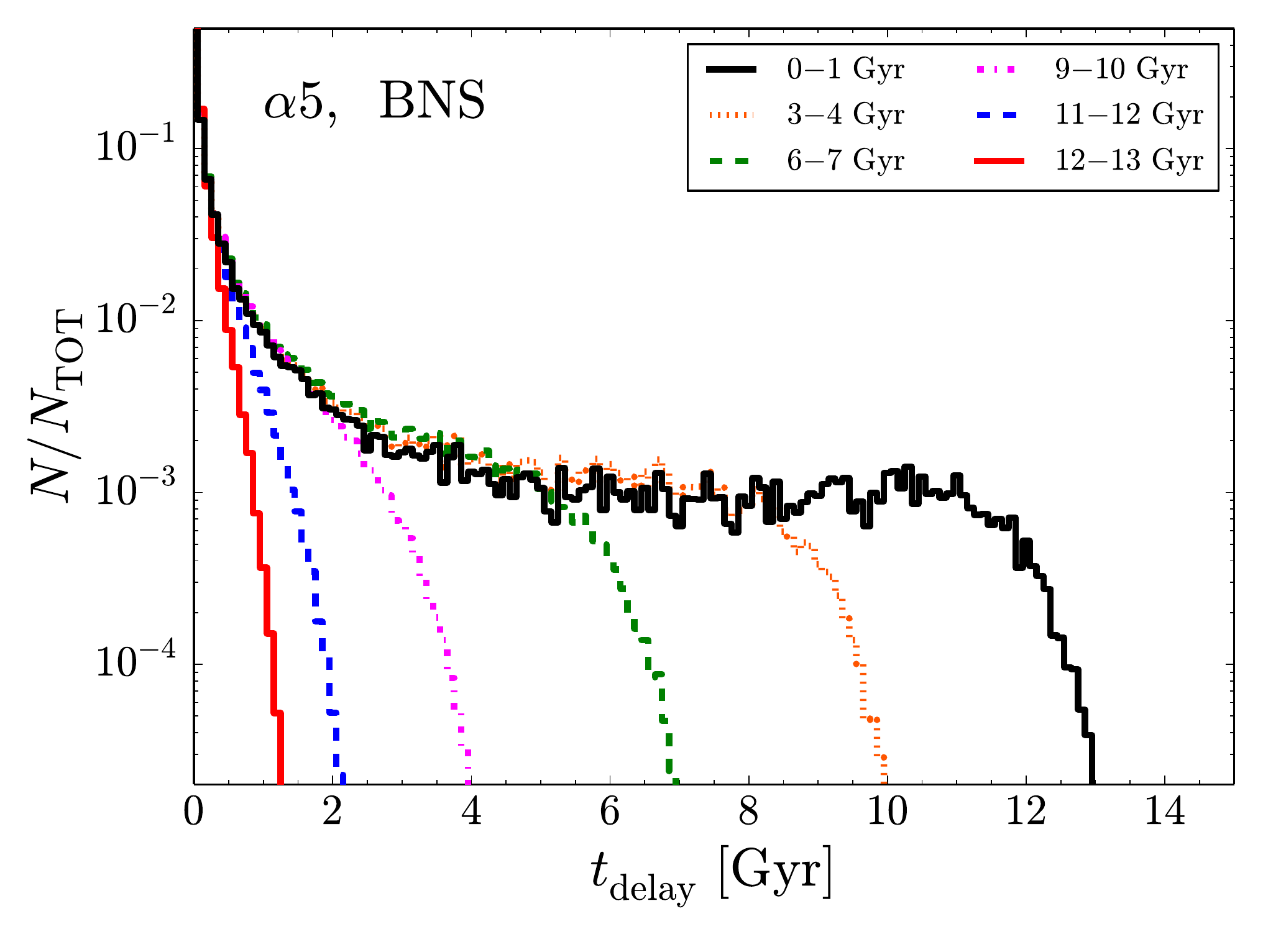,width=8cm} %was fig1.pdf
    \epsfig{figure=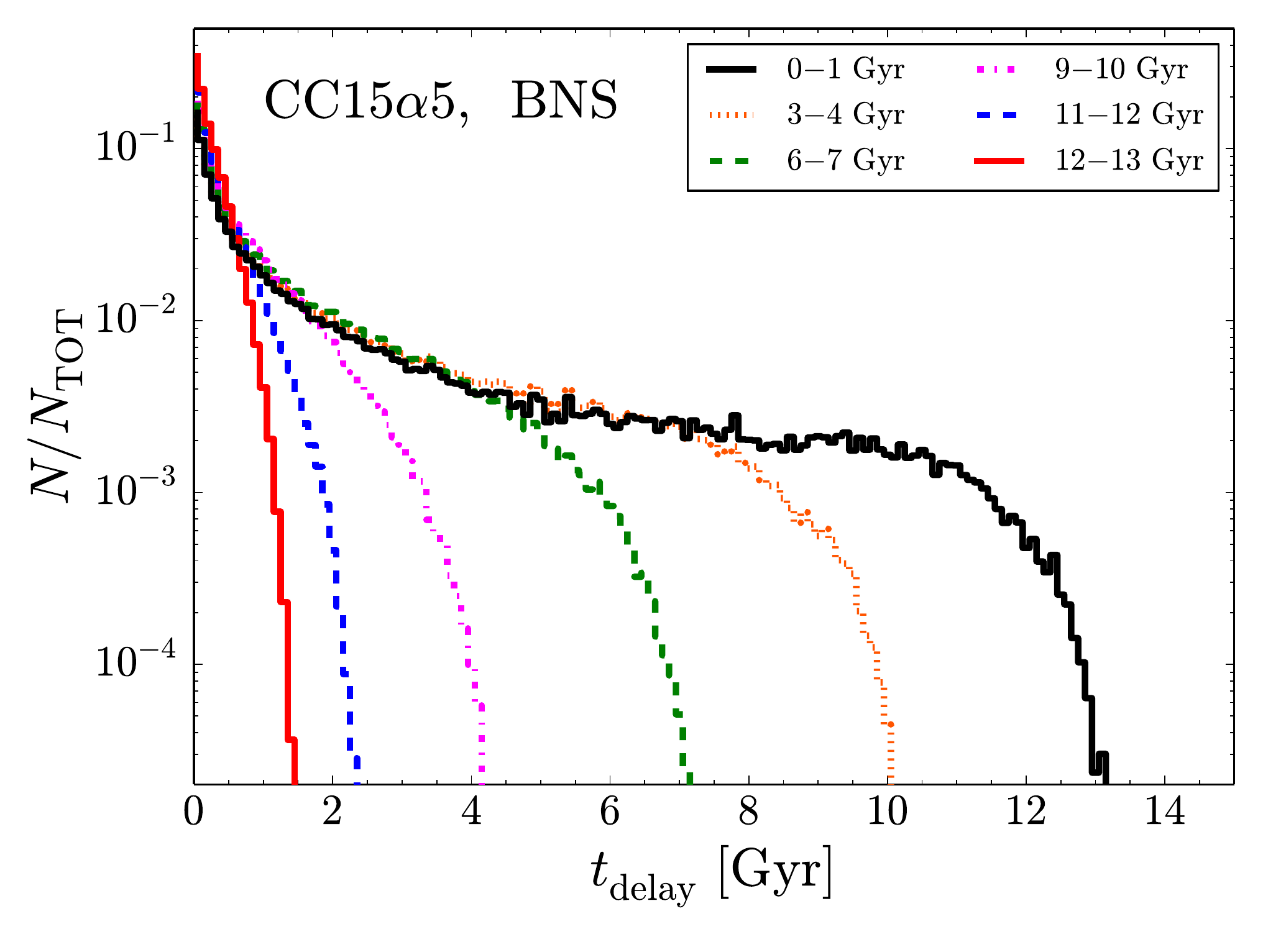,width=8cm} %was fig1.pdf
}}
\caption{Same as Fig.~\ref{fig:BBHdelay} but for BNSs.\label{fig:BNSdelay}
}
\end{figure}
%%%%%%%%%%%%%%%%%%%%%%%%%%%%%%%%%%%%%%%%%%%%%%%%%%%%%%%%%%%%%%%%%%%%%%%%%%%%%%%

%%%%%%%%%%%%%%%%%%%%%%%%%%%%%%%%%%% FIGURE 10 %%%%%%%%%%%%%%%%%%%%%%%%%%%%%%%%%%
\begin{figure}
\center{{
    \epsfig{figure=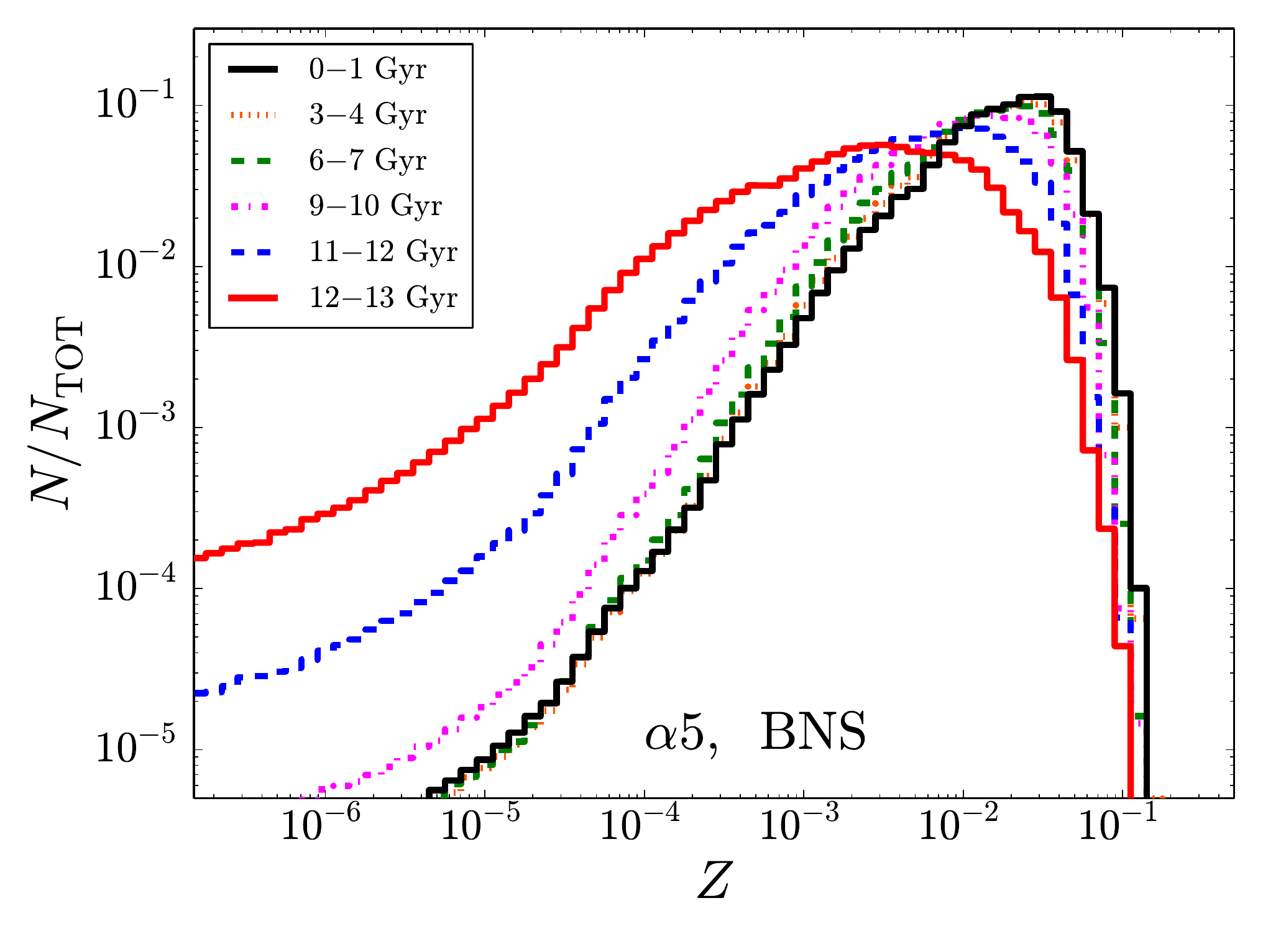,width=8cm} %was fig1.pdf
    \epsfig{figure=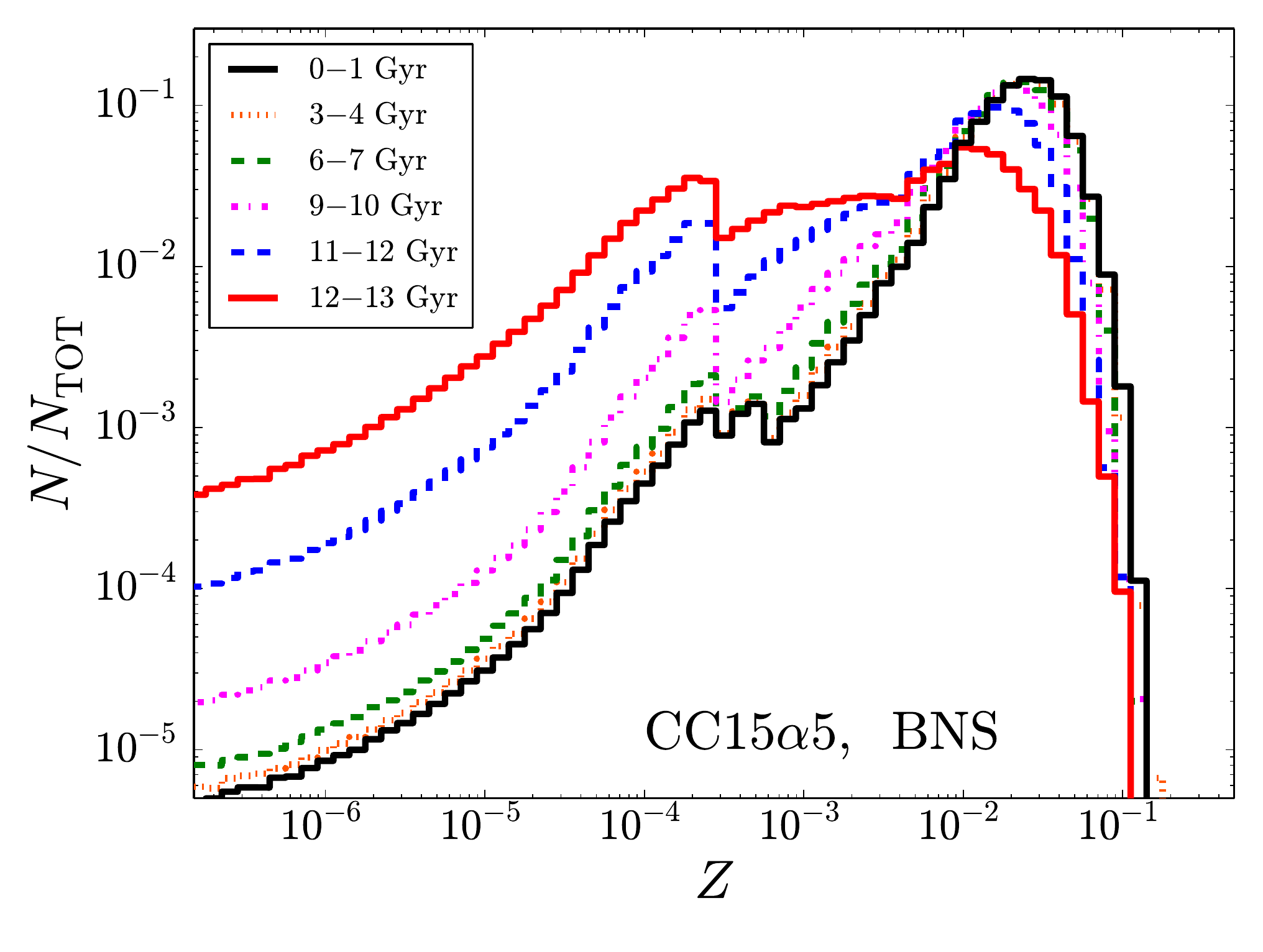,width=8cm} %was fig1.pdf
}}
\caption{Same as Fig.~\ref{fig:BBHZ} but for BNSs.\label{fig:BNSZ}
}
\end{figure}
%%%%%%%%%%%%%%%%%%%%%%%%%%%%%%%%%%%%%%%%%%%%%%%%%%%%%%%%%%%%%%%%%%%%%%%%%%%%%%%

Figure~\ref{fig:BNSmass} shows that even the mass spectrum of BNSs merging $12-13$ Gyr ago does not differ dramatically with respect to the mass spectrum of BNSs merging in the last Gyr. However, BNSs merging $12-13$ Gyr ago have a slight preference for larger masses than BNSs merging $<1$ Gyr ago. This is an effect of metallicity, as  we already discussed for the same population-synthesis runs in \cite{giacobbo2018b}: metal-poor binaries tend to produce slightly more massive NSs than metal-rich ones.

Differently from what happens to BHs (for which the BH mass dependence on metallicity does not translate into a BH mass dependence on merger redshift, because BBH mergers are orders of magnitude more common in a metal-poor population than in a metal-rich population),  in the case of BNSs the NS mass dependence with progenitor's metallicity translates also into a NS mass dependence with merger redshift, because the number of BNS mergers per unit stellar mass is approximately the same at high and low metallicity.

%*in contrast to NSs in NSBHs NSs in NSNS tend to be light
In contrast to NSBHs, NSs in merging BNSs tend to be light\footnote{As we already discussed in \cite{mapelli2018}, the NS mass distribution we derived adopting the prescriptions in \cite{fryer2012} tends to underestimate the typical mass of NSs by $\sim{}0.1$ M$_\odot$ with respect to observed Galactic NSs.} ($m_{\rm NS}<1.3$ M$_\odot$).

%* peak at solar metallicity - basically do not depend on Z then reflect what is the most common Z in the simulated Universe

%* tend to have much shorter merging time than both BBH and NSBHs. this is only partially  intrinsic difference (see figure by mapelli) and comes mostly from the metallicity dependence

Merging BNSs tend to have much shorter delay time than both BBHs and NSBHs (Figure~\ref{fig:BNSdelay}). This difference comes mostly from the metallicity dependence of the merger rate density of NSBHs and BBHs \citep{dominik2013,dominik2015,mapelli2018b}.

Even if the delay time of BNSs tends to be shorter than that of BBHs and NSBHs, we stress that $\sim{}17$\% and $\sim{}44$\% of all BNSs merging in the last Gyr have a delay time longer than 1 Gyr in runs~$\alpha{}5$ and CC15$\alpha{}5$, respectively (Table~\ref{tab:table2}, note that delay times of BNSs are significantly longer in run CC15$\alpha{}5$). This is a fundamental clue to understand why GW170817 is associated with an early-type galaxy: even if star formation is low in NGC~4993 nowadays, $\sim{}17-44$\% of all NSs which merge in the local Universe have formed $>1$ Gyr ago and are now locked in non-star forming massive galaxies.
%formed more than 6 Gyr ago. This is a fundamental clue to understand why GW170817 is associated with an early-type galaxy: even if star formation is low in NGC~4993 nowadays, more than ... \% of all NSs which merge in the local Universe have formed $>6$ Gyr ago and are now locked in non-star forming massive galaxies.

The metallicity of BNS progenitors (Figure~\ref{fig:BNSZ}) peaks at solar or super-solar metallicity: $\sim{}2-4\times{}10^{-2}$ for BNSs merging $0-1$ Gyr  ago. Figure~\ref{fig:BNSZ} basically reflects what is the most common metallicity in the Illustris simulation, because the number of BNS mergers per unit stellar mass does not significantly depend on metallicity. The only significant difference between $\alpha{}5$ and CC$15\alpha{}5$ is the secondary peak at $Z\sim{}10^{-4}$ in the latter simulation. This is an effect of the  dearth of BNS mergers from stars with metallicity $Z\sim{}4\times{}10^{-4}-4\times{}10^{-3}$ in the population synthesis simulations (see Figure~14 of \citealt{giacobbo2018b}).

\section{Discussion}\label{sec:caveats}
The main prediction of this paper is that the mass of merging compact objects does not depend (or depends mildly) on their merger redshift.

This result depends on a number of assumptions. First, it depends on the strong dependence of BBH and NSBH merger efficiency on metallicity predicted by population-synthesis simulations. In particular, {\sc{mobse}} predicts that the number of BBH and NSBH mergers originating from a metal-poor ($Z\leq{}2\times{}10^{-3}$) population is $3-4$ orders of magnitude larger than the number of BBH and NSBH mergers we expect from a metal-rich population with the same initial total mass. Thus, the number of merging BBHs and NSBHs which form from metal-poor stars is tremendously larger than the number of merging BBHs and NSBHs which form from metal-rich progenitors, even in the local Universe. Most BBHs/NSBHs that merge in the local Universe are expected to come from metal-poor progenitors which formed in the early Universe and have a long delay time. This means that the population of merging BBHs and NSBHs does not evolve with redshift because we are always looking at merging BBHs and NSBHs which formed from metal-poor stars, no matter what merger redshift we are considering.

Other population-synthesis codes (e.g. \citealt{belczynski2010b,dominik2013,dominik2015,stevenson2017,chakrabarti2017,kruckow2018,spera2018}) have predicted a similar trend (i.e. BBH mergers are more common from metal-poor progenitors than from metal-rich progenitors), but possibly with different strength. For example, in the simulations performed with  {\sc{sevn}} \citep{spera2018}, the number of mergers per unit stellar mass  is ``only'' two orders of magnitude lower at solar metallicity than at low metallicity (with \textsc{mobse} we find a difference of at least three orders of magnitude).   %%Thus, it is extremely important to repeat our calculations with different population-synthesis models.
%viene dal modello mobse in cui c'è una forte dipendenza del merger rate dalla metallicity - può essere + moderato si veda sevn

%Another peculiar feature of \textsc{mobse} is the distribution of delay times, which is sensibly longer than the one obtained with e.g. \textsc{startrack} \citep{dominik2012}. The distribution of delay times is another crucial ingredient of our models, because our simulated merger rate in the local Universe is dominated by BBHs with long  delay time. 

A  further crucial ingredient is metallicity evolution across cosmic time, which is very uncertain (see e.g. \citealt{madaudickinson2014,maiolinomannucci2018}). The model of sub-grid physics adopted in the Illustris is known to produce a mass-metallicity relation \citep{genel2014,genel2016} which is sensibly steeper than the observed one (see the discussion in \citealt{vogelsberger2013} and \citealt{torrey2014}). Moreover, the simulated mass-metallicity relation does not show the observed turnover at high stellar mass ($\gtrsim{}10^{10}$ M$_\odot{}$). In \cite{mapelli2017}, we estimate that the impact on BBH merger rate of the difference between the Illustris mass-metallicity relation and the observational relation \citep{maiolino2008,mannucci2009} is of the order of 20 \%. However, even if metallicity does not affect significantly the merger rate, it might affect the properties of merging compact objects (e.g. the masses). %Thus, future studies should repeat our calculation with different cosmological simulations and data-driven evolutionary models, to check the impact of metallicity evolution on the results (Santoliquido et al. in preparation).

 Thus, in a follow-up paper we will investigate the impact of different population-synthesis prescriptions and different cosmological evolution models, to properly assess the impact of all of these ingredients on the redshift evolution of compact binaries (Santoliquido et al. in preparation).

%rifare con altre simulazioni cosmologiche e altri modelli evolutivi

Moreover, in this work we neglect the dynamical formation channel of merging compact objects. Dynamics of young star clusters \citep{ziosi2014,mapelli2016,dicarlo2019} and globular clusters \citep{portegieszwart2000,downing2010,rodriguez2016,askar2017,hong2018} is known to favour the merger of more massive BBHs and to speed up their delay time \citep{dicarlo2019}. Thus, dynamics might significantly affect the mass and delay time of BBHs with respect to isolated binary evolution.

%no dinamica

%primordial BHs
Finally, we have not considered primordial BHs, i.e. BHs formed from gravitational instabilities in the early Universe \citep{carr1974}. Their very existence and their mass range are highly uncertain \citep{carr2016}. Any deviation from our results which cannot be explained with uncertainties on population synthesis models, dynamical effects, star formation or metallicity evolution in the Universe might suggest a different formation channel for BHs than the stellar one.

\section{Summary}
We have investigated the redshift evolution of several properties of merging compact objects (BBHs, NSBHs and BNSs) by means of population synthesis simulations. We used our population-synthesis code \textsc{mobse}, which adopts updated stellar wind models and prescriptions for electron-capture, core-collapse and (pulsational) pair instability SNe \citep{giacobbo2018,giacobbo2018b,giacobbo2019}. We have considered two very different prescriptions for natal kicks of core-collapse SNe: compact objects receive a natal kick distributed according to a Maxwellian distribution with one-dimensional root mean square velocity $\sigma_{\rm CCSN}=265$ km s$^{-1}$ and  $\sigma_{\rm CCSN}=15$ km s$^{-1}$ in run~$\alpha{}5$ and ~CC15$\alpha{}5$, respectively (in both cases the kick of BHs is modulated by fallback). We combined the results of \textsc{mobse} with the outputs of the Illustris cosmological simulations \citep{vogelsberger2014a,vogelsberger2014b,nelson2015} by means of a Monte Carlo formalism \citep{mapelli2017,mapelli2018,mapelli2018b}. With this procedure, we can account for redshift evolution of star formation rate and metallicity.

We find that the mass distribution of merging compact objects (even BBHs) depends only mildly on merger redshift (Figures~\ref{fig:BBHmass}, \ref{fig:NSBHmass}, \ref{fig:BNSmass}). This happens because the merger rate of BBHs and NSBHs depends dramatically on metallicity  \citep{belczynski2010b,giacobbo2018b}: the entire population of merging BBHs and NSBHs across cosmic time is dominated by metal-poor progenitors. Even if metal-rich stars should be more common in the local Universe and BBHs formed from metal-rich progenitors tend to be less massive than BBHs formed from metal-poor progenitors, we do not see the signature of BBHs originating from metal-rich stars because they are outnumbered by BBHs originating from metal-poor stars, even in the local Universe.

The only significant difference between BBHs merging in the last Gyr and BBHs merging more than 11 Gyr ago is that there is an excess of merging BBHs with mass $>20$ M$_\odot$ in the former population with respect to the latter (Figure~\ref{fig:BBHmass}). This happens because massive BHs ($m_{\rm BH}\sim{}20-35$ M$_\odot$) have preferentially long delay times. Thus, even if they form preferentially in the early Universe, they merge mostly in the local Universe.  It is important to note that this difference appears only if the CE efficiency parameter is large ($\alpha\ge{}3$). For smaller values of $\alpha{}$ even the BBH mass distribution does not change appreciably with redshift (see Figure~\ref{fig:BBHcheck}).

The mass of BHs in BBHs spans from $\sim{}5$ to $\sim{}45$ M$_\odot$, while the mass of BHs in NSBHs is preferentially low: most BHs in NSBHs have mass $<10$ M$_\odot$, especially in run~CC$15\alpha{}5$. The mass of NSs in NSBHs is preferentially large ($>1.4$ M$_\odot$), while the mass of NSs in BNSs is preferentially small ($<1.3$ M$_\odot$). 

The delay time distribution of BBHs (Figure~\ref{fig:BBHdelay}) is significantly flatter than that of NSBHs (Figure~\ref{fig:NSBHdelay}) and especially BNSs (Figure~\ref{fig:BNSdelay}), in agreement with previous work \citep{dominik2013,dominik2015}.  This is a consequence of the maximum stellar radius of the progenitors, which is significantly larger in BBHs progenitors, suppressing the formation of very close BBH binaries (even after CE). Thus, this result might depend on the stellar evolution models adopted in \textsc{mobse}.

 The typical progenitor's metallicity of merging BBHs and NSBHs is well below solar and evolves mildly with redshift (Figures~\ref{fig:BBHZ} and \ref{fig:NSBHZ}). In contrast,  the typical progenitor's metallicity of BNSs is solar or super-solar (Figure~\ref{fig:BNSZ}). Since the number of BNS mergers per unit solar mass does not depend on metallicity significantly \citep{dominik2015,giacobbo2018b}, the metallicity evolution of BNS progenitors traces the average metallicity evolution of the Universe, at least in the cosmological simulation.

The main prediction of this paper is that the  mass of merging compact objects does not depend (or depends very mildly) on their merger redshift. It is worth noting that we find no dramatic differences between run~$\alpha{}5$ and CC15$\alpha{}5$, despite the very different assumption for natal kicks. In this work, we have neglected the effect of dynamics. Dynamics tends to favour the merger of the most massive BHs (e.g. \citealt{mapelli2016,dicarlo2019}) and might significantly affect our conclusions. Moreover, the evolution of metallicity in the Universe is highly uncertain \citep{maiolinomannucci2018}. This sums up to uncertainties on binary evolution \citep{spera2018}. Thus, it is tremendously important to perform follow-up studies with a different treatment of metallicity and binary evolution.

\section*{Acknowledgments}
We thank the anonymous referee for their critical reading of the manuscript. We thank Daniel Wysocki, Alessandro Ballone, Alessandro Bressan, Emanuele Ripamonti and Mario Spera for useful discussions. 
We warmly thank The Illustris team for making their simulations publicly available. Numerical calculations have been performed through a CINECA-INFN agreement and through a CINECA-INAF agreement, providing access to resources on GALILEO and MARCONI at CINECA.  
 MM  acknowledges financial support by the European Research Council for the ERC Consolidator grant DEMOBLACK, under contract no. 770017. MCA acknowledges financial support from the Austrian National Science Foundation through FWF stand-alone grant P31154-N27 `Unraveling merging neutron stars and black hole - neutron star binaries with population-synthesis simulations'. NG acknowledges financial support from Fondazione Ing. Aldo Gini and thanks the Institute for Astrophysics and Particle Physics of the University of Innsbruck for hosting him during the preparation of this paper. This work benefited from support by the International Space Science Institute (ISSI), Bern, Switzerland,  through its International Team programme ref. no. 393 {\it The Evolution of Rich Stellar Populations \& BH Binaries} (2017-18).
 
\bibliography{./bibliography}
%\onecolumn
\appendix
\section{Dependence of BBH mass on CE efficiency and natal kicks}
         The two simulations presented in the main text (fiducial simulations) assume CE parameter $\alpha{}=5$. We have chosen $\alpha{}=5$ as a fiducial value, because Figure~15 of \cite{giacobbo2018b} shows that the BNS merger rate inferred from GW170817 \citep{abbottO2} is difficult to match  if we assume a lower value of $\alpha{}$. The choice of a different $\alpha{}$ does not affect significantly the masses of BNSs and NSBHs (see e.g. Figures~3 and 8 of \citealt{giacobbo2018b}), but might be important for BBHs (Figure~12 of \citealt{giacobbo2018b}). Moreover, the natal kicks of BBHs are quite similar in run~$\alpha{}5$ and ~CC15$\alpha{}5$, because these runs assume the same dependence on fallback.

          Thus, in this Appendix, we consider three additional models with different values of $\alpha{}$ and a different kick prescription, to check their impact on BBH masses. These additional runs are described in Table~\ref{tab:tableA1}. In particular, runs~$\alpha{}1$ and $\alpha{}3$ are the same as presented in \cite{giacobbo2018b}: they differ from run~$\alpha{}5$ only because $\alpha{}=1,\,{}3$, respectively. Run~K was already discussed in \cite{mapelli2017}. In this run, all BHs receive a natal kick randomly drawn from a Maxwellian distribution with $\sigma{}_{\rm CCSN}=265$ km s$^{-1}$. The kick is not reduced by the amount of fallback. Moreover, this run adopts the delayed SN model \citep{fryer2012}.  
 %%%%%%%%%%%%%%%%%%%%%%%%%%%%%%%%%% TABLE 1%%%%%%%%%%%%%%%%%%%%%%%%%%%%%%%%%%%%%
\begin{table}
\begin{center}
\caption{\label{tab:tableA1}
Properties of the supplementary population-synthesis simulations.}
 \leavevmode
\begin{tabular}[!h]{llllll}
\hline
Run    & $\alpha{}$ & $\sigma_{\rm ECSN}$ & $\sigma_{\rm CCSN}$ & Fallback & CCSN\\
& & [km s$^{-1}$] & [km s$^{-1}$] & & model \\
\hline
$\alpha{}1$     & 1.0        & 15  & 265  & yes & rapid\\
$\alpha{}3$     & 3.0        & 15  & 15   & yes & rapid\\
K               & 1.0        & 265 & 265  & no  & delayed\\ 
\hline
\end{tabular}
\begin{flushleft}
\footnotesize{Column 1: model name; column 2: value of $\alpha{}$ in the CE formalism; column 3 and 4: one-dimensional root-mean square of the Maxwellian distribution for electron-capture SN kicks ($\sigma_{\rm ECSN}$) and core-collapse SN kicks ($\sigma_{\rm CCSN}$). Column~5: yes (no) means that natal kicks are (are not) reduced by the amount of fallback. Column~6: core-collapse SN model (rapid or delayed from \citealt{fryer2012}). }
\end{flushleft}
\end{center}
\end{table}
%%%%%%%%%%%%%%%%%%%%%%%%%%%%%%%%%%%%%%%%%%%%%%%%%%%%%%%%%%%%%%%%%%%%%%%%%%%%%%%%
%%%%%%%%%%%%%%%%%%%%%%%%%%%%%%%%%%% FIGURE 1 %%%%%%%%%%%%%%%%%%%%%%%%%%%%%%%%%%
\begin{figure}
\center{{
    \epsfig{figure=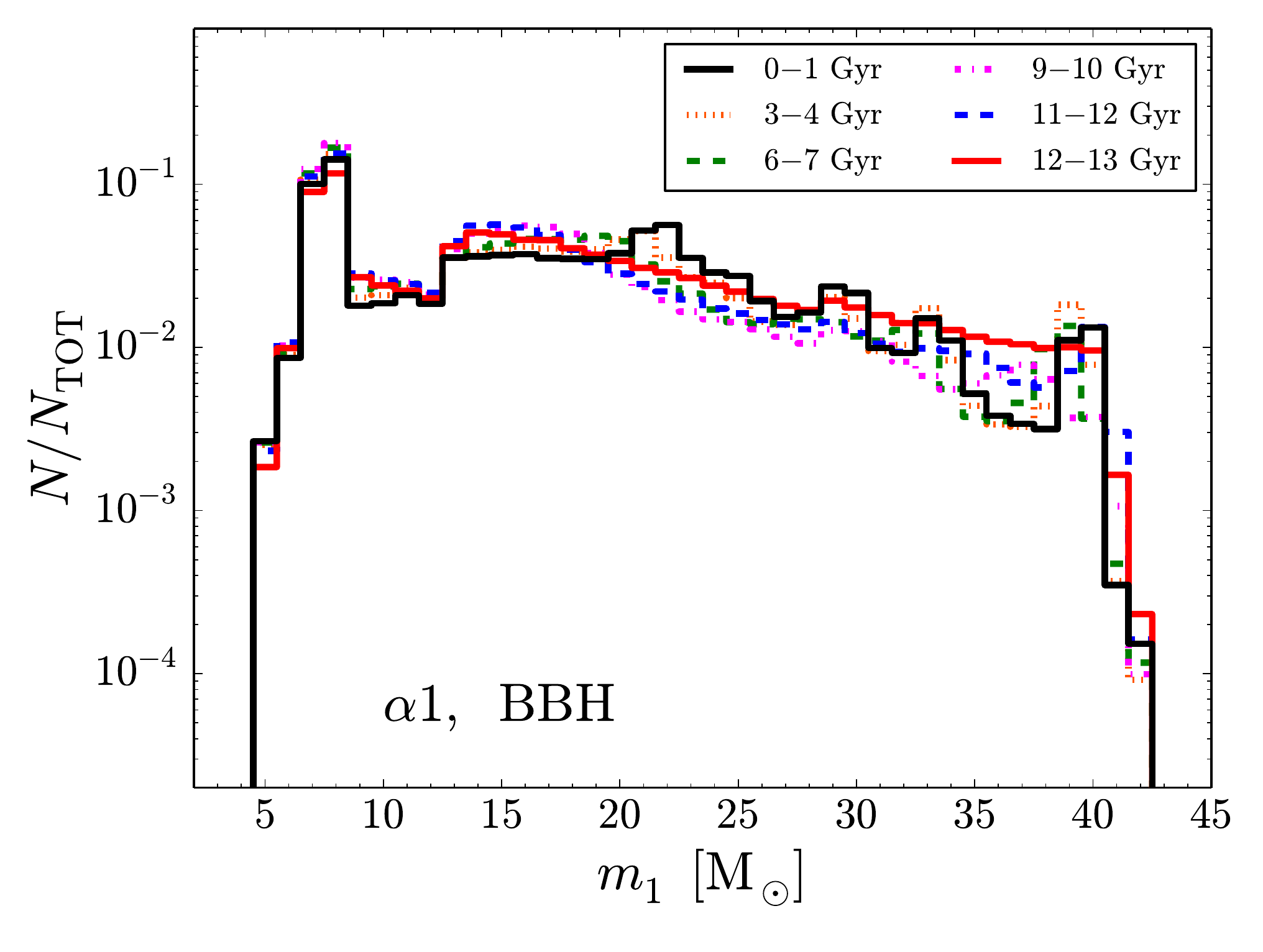,width=7cm} %was fig1.pdf
        \epsfig{figure=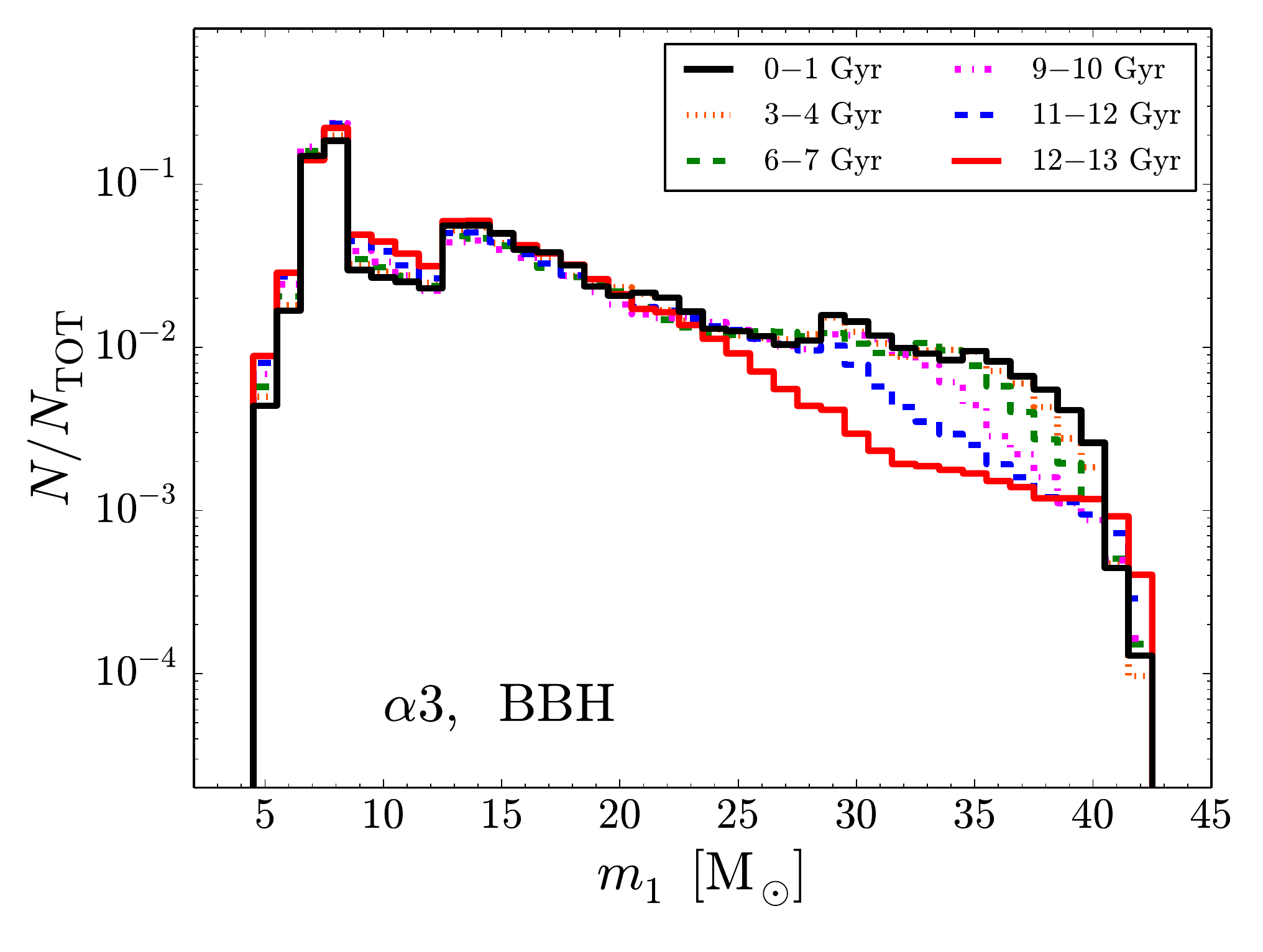,width=7cm} %was fig1.pdf
    \epsfig{figure=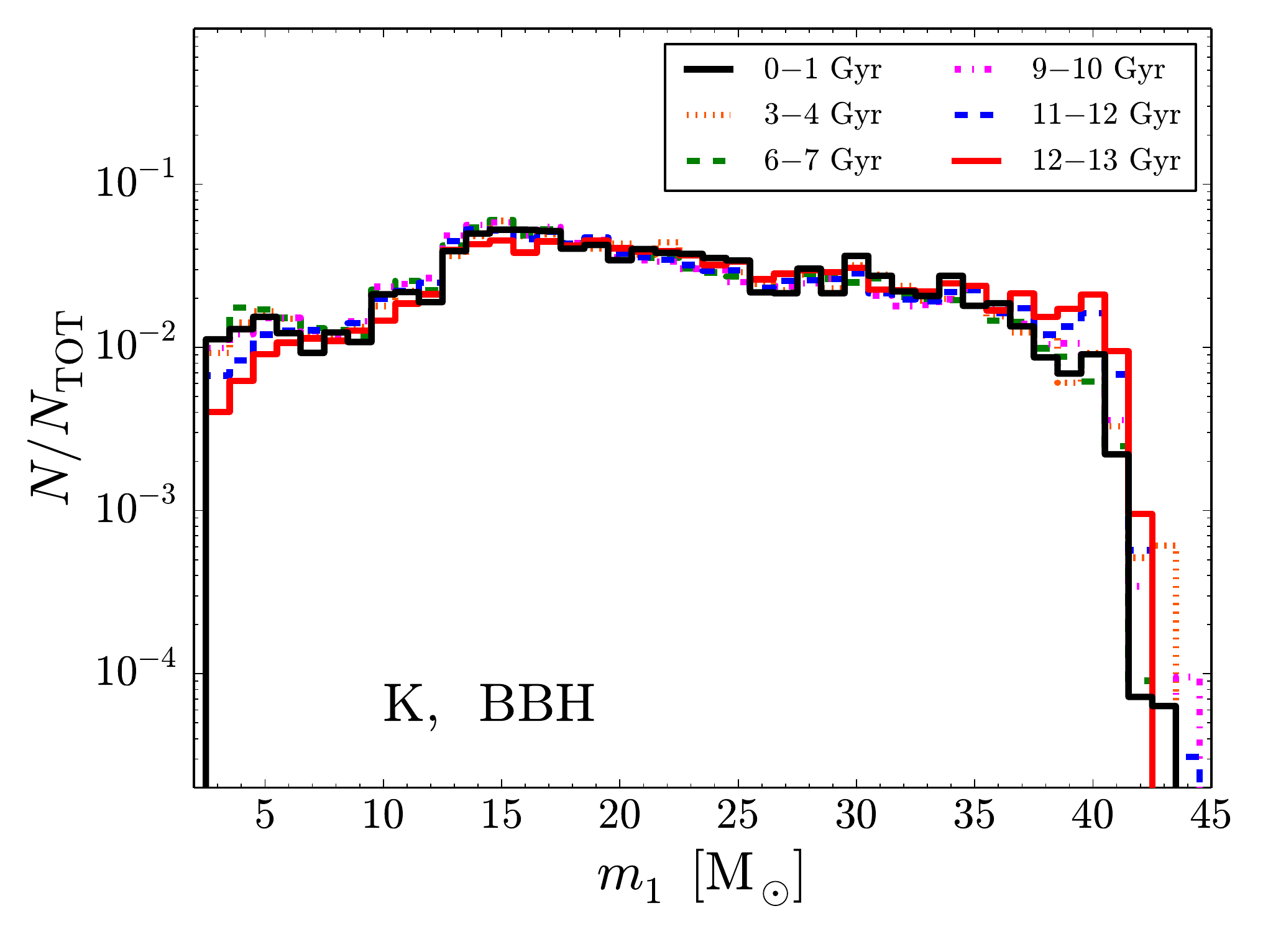,width=7cm} %was fig1.pdf
}}
    \caption{Mass of the primary BH ($m_1$)  in merging BBHs for the three supplementary runs. Top: run $\alpha{}1$. Middle: run~$\alpha{}3$. Bottom: run~K (from \citealt{mapelli2017}). 
      \label{fig:BBHcheck}
}
\end{figure}
%%%%%%%%%%%%%%%%%%%%%%%%%%%%%%%%%%%%%%%%%%%%%%%%%%%%%%%%%%%%%%%%%%%%%%%%%%%%%%%

 Figure~\ref{fig:BBHcheck} shows the distribution of primary BH masses in these three check runs. BBH masses vary very mildly with redshift even in these three runs. If $\alpha{}\le{}1$, even the slight excess of BHs with mass $\sim{}20-35$ M$_\odot$ merging in the last Gyr disappears: the mass of BHs does not evolve significantly with redshift.

\end{document}